\documentclass[10pt,twocolumn,cleanhead,cleanfoot]{config/asme2e}

\usepackage{empheq, amssymb}

\usepackage{lmodern}
\usepackage{amsmath}
\usepackage{amsfonts}
\usepackage{amssymb}

\usepackage{bm} % bold for greek symbols

\usepackage{txfonts}

\usepackage[T1]{fontenc}
\usepackage[utf8]{inputenc} % accept different input encodings

\usepackage[dvipsnames]{xcolor} % defines colors

\usepackage[american]{babel} %  culturally-determined typographical rules

\usepackage{graphicx}

\usepackage{epstopdf}

\usepackage[noadjust]{cite} % grouped citations

\usepackage{ltxcmds}

\usepackage{mathtools}

\usepackage{subcaption}% for subfigures

\usepackage{multirow} %for tables 

\usepackage{blindtext}

\usepackage{hyperref}

\usepackage{enumitem}

\usepackage{ifthen}

\usepackage{etoolbox}

\usepackage{cancel}

\usepackage{empheq}

\usepackage{fontawesome}

\usepackage{soul}

\usepackage{flushend}

\usepackage{textcomp}

\usepackage{lipsum}

\definecolor{light-gray}{gray}{0.4}
\definecolor{box-gray}{gray}{1}

\hypersetup{
    unicode=false,          % non-Latin characters in Acrobat’s bookmarks
    pdftoolbar=true,        % show Acrobat’s toolbar?
    pdfmenubar=true,        % show Acrobat’s menu?
    pdffitwindow=false,     % window fit to page when opened
    pdfstartview={FitV},    % fits the width of the page to the window
    pdftitle={Comparison of Data-Driven Modeling Approaches for Control Optimization of Floating Offshore Wind Turbines},    % title
    pdfauthor={Athul K. Sundarrajan and Daniel R. Herber},     % author
    pdfsubject={},   % subject of the document
    pdfkeywords = {}, % list of keywords
    pdfnewwindow=true,      % links in new window
    colorlinks=true,
    allcolors=blue
}

\usepackage{nomencl}
\newcommand{\xrm}[1]{\textrm{#1}}
\renewcommand\nomgroup[1]{%
  \item[\bfseries
  \ifstrequal{#1}{V}{ Variables}{%
  \ifstrequal{#1}{B}{ Subscripts}{%
  \ifstrequal{#1}{P}{ Notation}{%
  \ifstrequal{#1}{A}{ Acronyms}{}}}}]
}

\makenomenclature



\usepackage{dblfloatfix}
\usepackage{float}

\definecolor{block-gray}{gray}{0.95}

\usepackage[framemethod=TikZ]{mdframed}

\usepackage{xpatch}

\makeatletter
\xpatchcmd{\endmdframed}
  {\aftergroup\endmdf@trivlist\color@endgroup}
  {\endmdf@trivlist\color@endgroup\@doendpe}
  {}{}
\makeatother

\usepackage{fontawesome}

\usepackage{titlesec}

\usepackage{hyperref}

\newcommand{\doi}[1]{{doi:~\href{http://doi.org/#1}{#1}}\rmFullStop}

\newcommand*{\rmFullStop}{\rmifnextchar{.}{}{}}

\makeatletter
\newcommand{\rmifnextchar}[3]{%
  \begingroup
  \ltx@LocToksA{\endgroup#2}%
  \ltx@LocToksB{\endgroup#3}%
  \ltx@ifnextchar{#1}{%
    \def\next{\the\ltx@LocToksA}%
    \afterassignment\next
    \let\scratch= %
  }{%
    \the\ltx@LocToksB
  }%
}
\makeatother

\usepackage{colortbl}

\definecolor{light-gray}{gray}{0.6}

\newcommand{\xsection}[1]{\section[#1]{\MakeUppercase{#1}}}

\usepackage{algorithm2e}
\usepackage{xcolor}

\SetAlFnt{\footnotesize}

\SetAlCapSty{xAlCapSty}

\definecolor{commentcolor}{HTML}{1E4D2B}

\SetCommentSty{xCommentSty}

\SetNlSty{mynlfont}{}{} 

\LinesNumbered

\SetSideCommentRight

\DontPrintSemicolon

\RestyleAlgo{algoruled}

\usepackage{algorithm2e}
\usepackage{etoolbox}
\usepackage{xstring}
\usepackage{calc}
 
\newlength{\xalgowidth}
\newlength{\xalgoremainder}
\newlength{\xindentwidth}

\newenvironment{vAlgorithm*}[3][]{% before
  \setlength{\xalgowidth}{#2} % set algorithm width from second input
  \setlength{\xindentwidth}{#3} % set indent width from third input
  \setlength{\xalgoremainder}{\textwidth-\xalgowidth} % calculate indent to center the float
  \SetCustomAlgoRuledWidth{\xalgowidth} % set the rule width
  \IncMargin{\xindentwidth}
  \begin{algorithm*}[#1]
}% end before
{% after
  \end{algorithm*} 
  \DecMargin{\xindentwidth}
}% end after

{% after
  \end{algorithm} % column
  \DecMargin{\xindentwidth}
}% end after

\makeatletter
\patchcmd{\@algocf@start}{%
\begin{lrbox}{\algocf@algobox}%
}{%
\rule{0.5\xalgoremainder}{\z@}% indent
\begin{lrbox}{\algocf@algobox}%
\begin{minipage}{\xalgowidth}%
}{}{}
\patchcmd{\@algocf@finish}{%
\end{lrbox}%
}{%
\end{minipage}%
\end{lrbox}%
}{}{}
\makeatother

\usepackage{accents}

\definecolor{needcolor}{HTML}{C62828}

\newcommand{\DEL}{\ensuremath{\textrm{DEL}_t}}

\title{Data-Driven Modeling Approaches for Optimal Control and Control Co-Design of Floating Offshore Wind Turbines}
\author{Athul Krishna Sundarrajan\thanks{Corresponding author, \texttt{\href{mailto:atsun@dtu.dk}{atsun@dtu.dk}}}%
\affiliation{%
Postdoctoral Researcher \\
Department of Wind and Energy Systems \\
Danmarks Tekniske Universitet \\
Frederiksborgvej 399 \\
4000 Roskilde\\
\texttt{\href{mailto:atsun@dtu.dk}{atsun@dtu.dk}}}%
}

\author{
Daniel~R.~Herber%
\affiliation{%
Associate Professor \\
Department of Systems Engineering \\
Colorado State University \\
Fort Collins, CO 80523 \\
\texttt{\href{mailto:daniel.herber@colostate.edu}{daniel.herber@colostate.edu}}}%
}

\usepackage{amsmath}

\usepackage{microtype}

\begin{document}
 \setlength{\parskip}{0pt}
 \setlength{\parsep}{0pt}
 \setlength{\headsep}{0pt}

\setlength{\topsep}{0pt}
%\setlength{\partopsep}{0pt}

% equations
\abovedisplayshortskip=3pt
\belowdisplayshortskip=3pt
\abovedisplayskip=3pt
\belowdisplayskip=3pt

\titlespacing*{\section}{0pt}{18pt plus 1pt minus 1pt}{3pt plus 0.5pt minus 0.5pt}

\titlespacing*{\subsection}{0pt}{9pt plus 1pt minus 0.5pt}{1pt plus 0.5pt minus 0.5pt}

\titlespacing*{\subsubsection}{0pt}{9pt plus 1pt minus 0.5pt}{1pt plus 0.5pt minus 0.5pt}

\microtypesetup{nopatch=item}
\maketitle
\microtypesetup{patch=item}

\maketitle

%---------------
\begin{abstract}\noindent%
\textit{Models that balance accuracy against computational costs are advantageous when designing wind turbines with optimization studies, as several hundred predictive function evaluations might be necessary to identify the optimal solution.
We explore different approaches to construct low-fidelity models that can be used to approximate dynamic quantities and be used as surrogates for design optimization studies and other use cases.
In particular, low-fidelity modeling approaches using classical systems identification and deep learning approaches are considered against derivative function surrogate models ({DFSMs}), or approximate models of the state derivative function.
This work proposes a novel method that utilizes a linear parameter varying (LPV) modeling scheme to construct the DFSM.
We compare the trade-offs between these different models and explore the efficacy of the proposed DFSM approach in approximating wind turbine performance and different design optimization studies.
Results show that the proposed DFSM approach balances computational time and model accuracy better than the system identification and deep learning-based models.
Additionally, the DFSM provides nearly a fifty times speed-up compared to the high-fidelity model, while balancing accuracy.
%However, previous studies have assumed an a priori state dynamic model is available that can be directly evaluated to construct the DFSM.
%In this article, we propose an approach to extract the state derivative information from system simulations using piecewise polynomial approximations.
%Once the required information is available, we propose a multi-fidelity DFSM approach as a predictive model for the system's dynamic response.
%This multi-fidelity model consists of summation between a linear-fit lower-fidelity model and an additional nonlinear error corrective function that compensates for the error between the high-fidelity simulations and low-fidelity models.
%We validate the model by comparing the simulation results from the DFSM to the high-fidelity tools.
%The DFSM model is, on average, five times faster than the high-fidelity tools while capturing the key time domain and power spectral density~(PSD) trends.
%and use the DFSM for optimal control studies.
% The validation study shows that the multi-fidelity DFSM approach has comparable accuracy to traditional DFSM approaches and requires less time to construct.
%Then, an optimal control study using the DFSM is conducted with outcomes showing that the DFSM approach can be used for complex systems like floating offshore wind turbines~(FOWTs) and help identify control trends and trade-offs.%
}%
\end{abstract}%

\vspace{1ex}

%---------------
% \clearpage
\xsection{Introduction}\label{sec:introduction}

Models that accurately capture the dynamics are needed to identify and understand system-level optimal designs\footnote{A preliminary version of this article has been published in the 51st Design Automation Conference and is available at \doi{10.1115/DETC2025-167963}}. 
As a system's complexity and the fidelity of the underlying analyses increase, such models become more computationally expensive to evaluate~\cite{Deshmukh2017, Lefebvre2018}. 
For example, in highly-coupled multidisciplinary systems like floating offshore wind turbines (FOWTs) and marine turbines, the effect of the different subsystems (like the rotor, support structure (tower), floating platform, etc.) on each other must be fully considered to get an accurate response of the system~\cite{Pao2021}. 
Tools like OpenFAST~\cite{openFAST} and WEC-Sim~\cite{Ruehl2022} can accurately capture the dynamics of renewable energy systems like wind turbines, marine turbines, and wave energy converters.
A consequence of capturing the detailed dynamic response of these systems is the high computational cost associated with simulating them.
This computational expense, consequently, can make the direct use of these high-fidelity modeling tools in the following applications inefficient.

\subsubsection*{Estimating system performance:}~%
To evaluate the relevant performance metrics for systems like wind and marine turbines, the dynamic model must be simulated for several hundred design load cases~(DLCs) corresponding to different environmental conditions and operating scenarios as specified by the IEC design standards~\cite{IEC}.
Depending on the system, the simulation time for a single DLC can be several minutes to hours.
Therefore, a high computational cost in terms of CPU hours is incurred to evaluate the system's performance by simulating the system model over all relevant DLCs.

\subsubsection*{Use in design optimization studies:}~%
To identify the optimal physical design and/or control law, evaluating the dynamic system several hundred or more times is necessary.
Additionally, the software architecture of these system models might be such that it is impossible to link all the necessary variables of interest directly to an optimizer.
State variables are an example, as many simulation tools consider them internal, input-dependent quantities.
Engineers are typically interested in a variety of design optimization studies for different use cases, such as multi-objective studies to identify designs that balance trade-offs between different conflicting objectives, or control co-design (CCD) studies to simultaneously optimize for both control and plant variables associated with a system to identify system-level optimal designs.

To solve these problems efficiently, engineers often use gradient-based solvers.
Many simulation tools do not provide the accurate gradient information that these gradient-based solvers require.
One recourse is to use finite-difference approaches to estimate the gradients, but these gradient estimates for simulations can be noisy or insensitive to small perturbations in the design variables and can be expensive to evaluate.
For these reasons, engineers are limited when setting up design optimization studies and resort to using gradient-free solvers like genetic algorithms (GA), which require several hundred function evaluations, or methods like COBYLA, which do not scale well with the number of design variables~\cite{Zalkind2022}.

Therefore, computationally inexpensive models that can capture some key dynamic quantities like the system states and metrics with sufficient accuracy could be helpful~\cite{Azad2023, Deshmukh2017, Sundarrajan2023}.
These approximations are generally referred to in the literature as low-fidelity models because of their limited scope in capturing the system response.
The necessity of low-fidelity models for many applications with computationally expensive high-fidelity models has been outlined in detail in design optimization literature.
The scope of this article is limited to developing low-fidelity models for the open-source wind and marine turbine simulation tool OpenFAST.
Broadly, there are two different ways to construct low-fidelity models for wind turbines, namely physics-based and data-driven.
The focus of this article will be on constructing data-driven low-fidelity models of wind turbines that can address the two drawbacks of using high-fidelity models outlined above.
Section~\ref{sec:model-requirements} outlines the problem statement and the requirements of the low-fidelity model that is developed to address the drawbacks outlined above for OpenFAST.
A more detailed outline for the rest of the article is presented at the end of Sec.~\ref{sec:model-requirements}.

\xsection{Background}\label{sec:model-requirements}

\begin{figure*}[t]
    \centering
    \includegraphics[scale = 1]{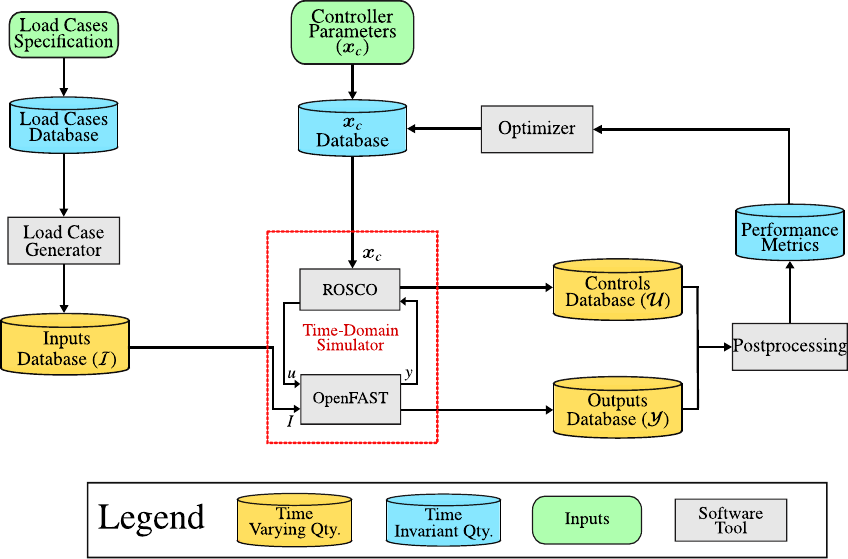}
    \caption{A high-level workflow diagram illustrating the different components associated with simulating and/or performing design optimization studies for a wind turbine system.}
    \label{fig:workflow}
\end{figure*}

A high-level workflow diagram illustrating the different steps and software components associated with simulating a wind or marine turbine model using OpenFAST is shown in Fig.~\ref{fig:workflow}. 
At the heart of this process are OpenFAST~\cite{Jonkman2016} and the open-source controller ROSCO introduced in Ref.~\cite{Abbas2021}.
To understand Fig.~\ref{fig:workflow} and the motivation behind this article, it is necessary to understand how these core components are set up and operate.

\subsection{OpenFAST}\label{subsec:openfast}

OpenFAST is an open-source modeling tool that is extensively used to simulate the dynamic response of fixed or floating wind turbine systems to various wind and wave excitations.
Although primarily used for wind turbines, OpenFAST has been recently extended to model the response of marine turbine systems as well~\cite{Murray2018, Wiley2023}.
%Marine turbines operate similarly to wind turbines.
%They convert the kinetic energy of flowing water to electrical energy.
Given a description of the different components, such as the rotor, generator, tower, platform mooring lines, etc., and their properties, OpenFAST can predict the aerodynamic and hydrodynamic loads, generator performance, and the displacement, velocities, and reaction loads of the turbine components, for the exciting wind and wave elevation signals.
OpenFAST can be broken down into a set of interconnected modules that contain the relevant physics needed to model the different aerodynamic, hydrodynamic, generator, and structural reactions~\cite{Jonkman2016}.
OpenFAST formulates the dynamics of the turbine as a set of differential-algebraic equations (DAEs) and solves these DAEs to predict the dynamics.
Because OpenFAST needs to resolve the underlying physics of the different modules individually and account for the coupling between modules, it can be computationally expensive.
%This is the case when simulating marine turbines using OpenFAST, where the average simulation time can be of the order of several hours for a single simulation.

\begin{figure}[t]
    \centering
    \includegraphics[scale = 1]{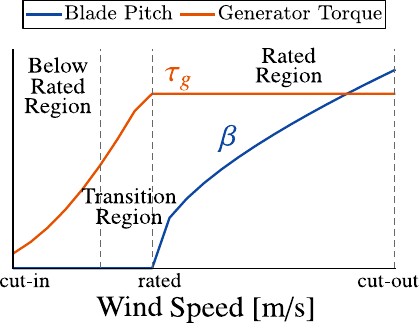}
    \caption{Generator torque and blade pitch vs. wind speed.}
    \label{fig:wind-reg}
\end{figure}

\subsection{Control of Wind Turbines}
\label{subsec:controls}

% new paragraph
Wind turbines must operate in stochastic offshore environments where they can be subject to extreme forces.
A controller is essential to mitigate the loads the turbines are exposed to and ensure power production~\cite{Pao2009, Abbas2021}.
Recently, several studies have investigated the importance of the controller and its performance in reducing the lifetime costs associated with wind and marine systems~\cite{Abbas2021, Sundarrajan2023, Zalkind2022, Jonkman2021, Ross2022}.
The wind turbine system under consideration in this study requires two control inputs: the generator torque~($\tau_g$) and the blade pitch~($\beta$).
Typically, the operating region between a cut-in and cut-out wind speed is subdivided into two regions: below-rated and rated.
Different control goals and variables are used in these regions.
In the below-rated region, $\tau_g$ is the primary control variable; in the rated region, $\beta$ is the primary control variable.
This behavior is illustrated in Fig.~\ref{fig:wind-reg}.

%In this article, we use the open-source reference controller ROSCO, along with OpenFAST to perform dynamic simulations.
Designing and tuning the controllers for wind and marine turbines can be time-consuming. 
Typically, the system models are linearized around set operating points, and the controllers for both $\tau_g$ and $\beta$ are designed through Bode shaping~\cite{Abbas2021}.
The controller gains are usually functions of specific parameters, denoted by $\bm{x}_c$, that can be tuned to improve the controller's performance.
Typically, these parameters were chosen by a control engineer.
ROSCO was designed to overcome these drawbacks and automate the controller design and tuning process~\cite{Abbas2021}. 
ROSCO achieves this by using generic models of the turbine, which are computationally inexpensive to linearize, and using optimizers to identify $\bm{x}_c$.
Two different control loops are implemented for $\tau_g$ and $\beta$ that use a proportional-integral (PI) architecture.
The main feedback variable for both these control loops is the generator speed ($\omega_g$) signal, which is obtained from OpenFAST.
The closed-loop system between $\omega_g$ and the control response $\tau_g$/$\beta$ is a second-order system whose response can be characterized by its natural frequency ($\omega$) and the damping ratio ($\zeta$).
For floating turbines, to address the `negative-damping problem', another proportional controller is utilized~\cite{Abbas2021, Veen2012}.
The tower-top acceleration ($\ddot{x}_t$) is the feedback variable for this controller, and is proportional feedback with a gain of $k_{\beta,\xrm{float}}$, and the output of this closed-loop system is added to the blade pitch response from the PI controller.

The controller's performance can be improved by selecting appropriate values of these parameters $\omega$, $\zeta$, and $k_{\beta,\xrm{float}}$.
Typically, for a given turbine model and a set of DLCs, an optimization problem is set up to find values of $\bm{x}_c = [\omega,\zeta,k_{\beta,\xrm{float}}]^\mathrm{T}$ that maximize or minimize specific performance metrics, which are calculated by simulating the turbine model using OpenFAST.
This process is shown in Fig.~\ref{fig:workflow}.
Solving this optimization problem using OpenFAST simulations can be challenging for the reasons outlined in Sec.~\ref{sec:introduction}.
Therefore, this article's key focus will be on constructing a low-fidelity model that can predict the performance metrics for expensive simulations and accelerate the design optimization process.
With the core elements of Fig.~\ref{fig:workflow} explained, Sec.~\ref{subsec:workflow-elements} will briefly outline the rest of the elements of Fig.~\ref{fig:workflow}.

\subsection{Process Workflow}\label{subsec:workflow-elements}

The process starts with the specification of the load cases, which characterize the wind speed and wave elevation signals, which act as key inputs ($\mathcal{I}$) to the simulation.
These specifications are usually the metocean (meteorological-oceanographic) conditions at the site where the turbine is located, which are usually quantities like the average wind speed ($\bar{w}$), the turbulence intensity of the wind speed signal, which is the ratio of the standard deviation to the mean, and the height and time period of the incoming waves.
These specifications are passed to the load case generator, which generates the time series of the wind/current speed and wave elevation signals.
The next step in the process is to carry out the simulations for these load cases.
OpenFAST passes the instantaneous values of key feedback variables like $\omega_g$ to the controller, and the controller calculates the instantaneous values of $\tau_g$ and $\beta$ back to OpenFAST.
The results of these simulations, which are the time series of the key input, control, and output signals, are stored and passed on to the postprocessing tool, which calculates key performance metrics from these tools.
These performance metrics are used as objectives and constraints in the optimization studies and are passed on to the optimizer.

\begin{figure*}[t]
\centering
\begin{subfigure}[t]{0.5\textwidth}
\centering
\noindent\includegraphics[scale=1]{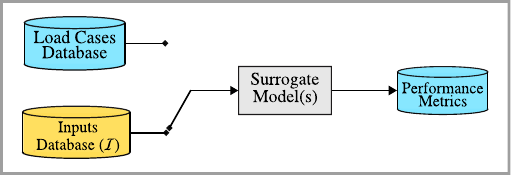}
\caption{Surrogate modeling approach 1.}
\label{fig:surrogate1} 
\end{subfigure}%
%---------------------------------------------------
\begin{subfigure}[t]{0.5\textwidth}
\centering
\noindent\includegraphics[scale=1]{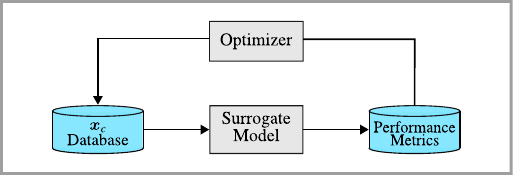}
\caption{Surrogate modeling approach 2.}
\label{fig:surrogate2} 
\end{subfigure}%
%------------------------------------------------
%-----------------------------------------------------------------
\caption{Two different surrogate modeling approaches that can be used for wind turbines. The first approach shown in Fig.~\ref{fig:surrogate1} can be used to train a surrogate model that predicts key performance metrics given information about the load cases. The second approach, shown in Fig.~\ref{fig:surrogate2}, has been used extensively to construct surrogate models that can be used for design optimization studies.}
\label{fig:surrogate-approaches}
\end{figure*}

\subsection{Surrogate Modeling Methodology}\label{subsec:surrogates}

% new paragraph
Different surrogate modeling approaches have been utilized to address the two drawbacks outlined in Sec.~\ref{sec:introduction}.
Using the elements presented in Fig.~\ref{fig:workflow}, these surrogate modeling approaches can be visualized as shown in Fig.~\ref{fig:surrogate-approaches}. 
The first approach to constructing surrogate models is to map key performance metrics like fatigue damage or AEP, given load case specifications like average wind speed, wave elevation, turbulence intensity, etc, or their time series~\cite{Haghi2024}.
This approach is shown in Fig.~\ref{fig:surrogate1}.
A variety of different data-fitting approaches (e.g.,~artificial neural networks, Gaussian process models, recurrent neural networks, etc.) have been used to construct these surrogate models.
Although some of these studies have used time-varying quantities as inputs to the surrogate models, the outputs predicted are time-invariant quantities.

The second set of approaches, which has been widely used in surrogate-assisted design optimization, involves constructing a surrogate model to predict the performance metrics given the design variables in question~\cite{Peherstorfer2018, March2012}.
This approach is shown in Fig.~\ref{fig:surrogate2}.
Typically, a set of load cases is identified, and a sampling scheme for the design variables is created.
Then, the system is simulated to get the values of the key performance metrics at these sample points.
A surrogate is then trained to predict the performance metrics given the design variables as the inputs, and is used in the design optimization process. 

\begin{figure*}[t]
\centering
\begin{subfigure}[t]{0.5\textwidth}
\centering
\noindent\includegraphics[scale=1]{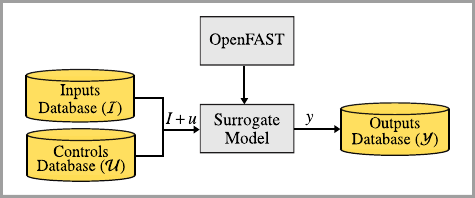}
\caption{Model construction.}
\label{fig:dfsm-construct} 
\end{subfigure}%
%---------------------------------------------------
\begin{subfigure}[t]{0.5\textwidth}
\centering
\noindent\includegraphics[scale=1]{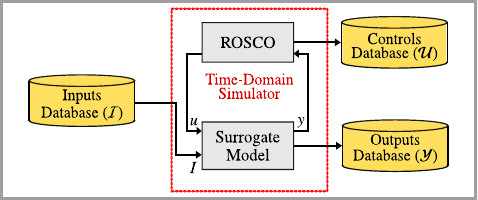}
\caption{Model usage.}
\label{fig:dfsm-usage} 
\end{subfigure}%
%------------------------------------------------
%-----------------------------------------------------------------
\caption{An alternate approach to construct a surrogate model that can be used for predicting time-series of key outputs, given inputs and controls.}
\label{fig:dfsm-constrcut-usage}
\end{figure*}

It is straightforward to combine these two surrogate modeling approaches to create a low-fidelity model that can be used for these two applications.
However, there are some drawbacks to using the approaches as they are presented in the aforementioned studies to construct the low-fidelity model.
These studies have presented different approaches to predict time-invariant quantities associated with wind turbines.
However, to understand the behavior of these systems, it is essential to understand the key trends in the dynamic quantities, represented as the controls and outputs database in Fig.~\ref{fig:workflow}, along with the performance metrics.
For example, in controller optimization studies, to understand why a set of $\bm{x}_c$ identified by the optimizer results in minimizing/maximizing the key performance metrics, it is necessary to look at the time series signals of key controls and outputs that $\bm{x}_c$ results in, which these surrogate models do not capture~\cite{Sundarrajan2024}.
Additionally, in early-stage design studies, engineers often run multiple preliminary studies to explore the effect of optimizing for different sets of design parameters.
When the set of design parameters changes, the process outlined above must be repeated to construct a new surrogate model, which can be computationally expensive and inefficient.

% new paragraph
Therefore, in this article, we explore different approaches to constructing surrogate or low-fidelity models that can be used to predict the time series of the control variables and key outputs.
Typically, in the workflow diagram presented in Fig.~\ref{fig:workflow}, evaluating OpenFAST is often the most computationally expensive operation.
As shown in Fig.~\ref{fig:workflow}, OpenFAST should be characterized as a high-fidelity model that predicts the key outputs, given inputs, and controls.
Therefore, constructing a low-fidelity model that can predict the key outputs and be coupled with the controller to perform closed-loop simulations would address the two problems outlined in Sec.~\ref{sec:introduction}.
This approach is visualized in Fig.~\ref{fig:dfsm-constrcut-usage} using the elements discussed in Fig.~\ref{fig:workflow}.
We limit our focus to testing these low-fidelity models for controller optimization studies.

The main requirements for the surrogate model can be summarized as follows:
\begin{enumerate}
    \item The user cannot access the underlying DAEs in OpenFAST directly, and can only interact with OpenFAST through simulations.
    When constructing surrogates, the model must, therefore, be built using time series data for key inputs, controls, and outputs.
    \item Once constructed, the model will be coupled with the controller, and it must be able to predict the time series response of key outputs, given inputs and controls.
\end{enumerate}

% new paragraph
The rest of the article is organized as follows.
In Sec.~\ref{sec:low-fidelity-model}, we discuss the different approaches that can be used to construct low-fidelity models, and identify three different approaches that can be used to construct such a model.
These are systems-identification approaches, deep learning, and derivative function surrogate models (DFSM).
In Sec.~\ref{sec:model-construct} we outline the novel approach to construct the DFSM model, and in Sec.~\ref{sec:validation} we construct low-fidelity models using these three approaches, compare their performance and trade-offs for a validation test, and identify the best approach.
In Sec.~\ref{sec:usecases}, we further test the identified model to see if it can be utilized for different types of design optimization studies, such as a multi-fidelity controller optimization study, and a multi-objective CCD study.
Finally, in Sec.~\ref{sec:conclusion}, we summarize the findings and provide directions for future research.

\xsection{Low-Fidelity Models}\label{sec:low-fidelity-model}

% \subsection{A Note on Semantics}

% % new paragraph
% The terms surrogate model, metamodel, and low-fidelity models are often used interchangeably and refer to a computationally inexpensive approximation of an expensive model.
% We use the different terms in the following context. Surrogate models can be any data-fit approach that maps inputs to outputs. Their scope is fixed, and can only be used to predict the set of outputs to the inputs.
% Low-fidelity models can be a combination of different surrogate models, such that their scope is larger than the individual surrogate models. They can be data-driven or physics-based, or a combination of both.
%  that fit the requirements outlined in Sec.~\ref{subsec:surrogates}, namely physics-based or data-driven.
Primarily, there are two broad approaches that can be used to construct low-fidelity models of wind turbines, namely physics-based and data-driven, which are now summarized.
 
\subsection{Physics-Based Models}

Physics-based approaches try to approximate the key relationships using prior knowledge of the underlying physics of the system.
Typically, this process starts by identifying simple relationships that approximate the behavior of key subsystems~\cite{Jonkman2022, Wang2017, Vercellino2022}. 
Then, parameters in these relationships (e.g., the stiffness constant of the force-displacement relationship) can be found using analytic expressions or data and system response outputs predicted for different inputs.
These simplified relationships can also be augmented to obtain more accurate ones that better represent the behavior of the system of interest.
This additional model accuracy can lead to detailed mathematical representations closely approximating reality but may be computationally expensive to evaluate~\cite{Bayat2025, Lee2025}.
These types of models are often referred to as white-box models, where the relationship between the quantities can be obtained by solving the relevant governing equations.
This is a bottom-up way to construct a physics-based low-fidelity model.

A top-down approach could also be used to construct physics-based low-fidelity models.
These approaches start from high-fidelity models and obtain different types of reduced-order or low-fidelity models.
There are several ways to obtain these reduced-order models, depending on the high-fidelity model.
For high-fidelity models such as OpenFAST, using Taylor-series approximations to obtain linearized model(s) is a widely used approach to obtain low-fidelity models~\cite{Sundarrajan2023}.
Another approach is to use simplifying assumptions of the system behavior/property that enable simple models~\cite{Peherstorfer2018}.
For example, in OpenFAST, assuming that the rotor blades and tower can be represented as Euler-Bernoulli beams with limited deflections allows users to use a simplified solver to obtain the corresponding states associated with these subsystems~\cite{Wang2017}.
Furthermore, developing these physics-based low-fidelity models can be time-consuming and requires technical expertise and background regarding the fundamental physics of the system.
Instead, in this article, we focus on using data-driven approaches that can be used to develop multi-input multi-output low-fidelity models.

% new paragraph
\subsection{Data-Driven Models}

Data-driven low-fidelity models capture the input-output relations using different approaches like curve-fitting, machine learning, probabilistic models, etc.~\cite {Haghi2024}.
Unlike physics-based low-fidelity models, which can be complex to develop and construct, data-driven models can be straightforward to build.
Because they map input-output relations, data-driven approaches can be flexible and are used in a wide variety of scenarios and applications~\cite{Zhao2022}.
This facet of data-driven models allows them to capture highly nonlinear and layered relationships between quantities, which may be difficult using physics-based models.
These data-driven approaches have traditionally been used to capture the input-output relationship between time-invariant quantities.
However, with the advent of deep learning techniques like recurrent neural networks (RNNs) and long short-term memory (LSTM) networks, data-driven approaches have also been used to capture the relationship and trends between dynamic quantities~\cite{SiamiNamini2019}.

In addition to machine learning approaches like the ones mentioned above, other approaches can be used to build low-fidelity/surrogate models of dynamic systems. 
Systems identification approaches have been widely utilized to construct low-fidelity models for various theoretical and real-world dynamic systems, and these approaches are built on rigorous theory~\cite{Kollar1997, Overschee1996, Kerschen2006, Worden2018}.
At this point, we would like to categorize data-driven approaches into those that assign a model structure to the input-output relations and those that do not.
Such a distinction has also been made in Ref.~\cite{Zhao2022}.
This distinction is helpful as it allows us to discuss specific modeling approaches from the field of systems identification as data-driven modeling approaches.
For example, approaches like LSTMs do not assume a model/relationship structure between the inputs and outputs.
In contrast, there are many approaches used in systems identification that often assume that the inputs and outputs can be related through a mathematical model.
By specifying a model structure, the systems identification approaches can also guarantee certain properties about the identified model as well~\cite{Ljung1998, Jansson2005}.
For example, it is possible to ensure that the system identified is stable, which is a key requirement when using these models for closed-loop control simulations.
%It is difficult to make such guarantees with approaches like LSTMs.
%In this article, we will explore the benefits/trade-offs of both model-based and model-free data-driven modeling approaches for wind and marine turbines.

Model-based system identification approaches can further be broadly classified into two categories depending on how the model is constructed, namely parameter estimation methods and subspace identification methods.
Parameter estimation methods assume that the inputs and outputs are related through a system model~($\bm{M}$) that has unknown parameters~$\bm{M}(\bm{\theta})$~\cite{Ljung1998}.
Parameter estimation methods then find the optimal $\bm{\theta}^{*}$ that minimizes the error between the actual model outputs and the outputs predicted by $\bm{M}(\bm{\theta}^{*})$.
On the other hand, subspace identification approaches identify a discrete linear time-invariant (LTI) system that relates how the outputs evolve with the inputs~\cite{Overschee1996}.
The linear system states and the corresponding matrices are identified by applying singular value decomposition to the Hankel matrix of the time series data.
Unlike parameter estimation methods, subspace identification methods are not iterative.

Other model-based data-driven approaches have also been successfully used to construct surrogate models for wind turbines.
These approaches can be generally termed as derivative function surrogate modeling (DFSM) approaches investigated in Refs.~\cite{Deshmukh2017, Lefebvre2018, Zhang2022, Qiao2021}.
DFSM captures the relationship between the inputs $(\bm{u})$ and the system states $(\bm{\xi})$ and assumes they are related through derivative function or dynamic model $\bm{f}(\cdot)$ that describes the first-time derivatives of the states $(\dot{\bm{\xi}})$ such that $\dot{\bm{\xi}} = \bm{f}(\bm{u},\bm{\xi})$.
The DFSM is a surrogate of $\bm{f}_{\textrm{DFSM}}(\cdot) \approx \bm{f}(\cdot)$ that predicts how the state derivatives evolve with respect to the controls and states.
This approach was developed to be used for continuous-time open-loop optimal control studies, as popular solution methods for these problems require information about how the state derivatives evolve over time for different controls and states~\cite{Herber2020d, Allison2014}.

% new paragraph
The assumption that a state-space model can effectively approximate the system behavior is common in DFSM and system-identification approaches.
However, because of how these models have been constructed and utilized for different applications, previous studies using the DFSM approach have not tried to understand it from a systems identification perspective.
Instead, DFSM has been primarily understood from a surrogate modeling perspective.
System-identification approaches typically build models using time series data of the key inputs and outputs of the system.
Once constructed, these models have been used for applications such as model prediction, control development, etc.
In contrast, the previous instances of the DFSM approach assume that the state-derivative function is available in a form that can be sampled for different values of the controls and states to get the corresponding state-derivative values~\cite{Zhang2022, Deshmukh2017, Lefebvre2018}.
Subsequently, a data-fitting method, like radial-basis functions (RBF) or GPR, is used to construct $\bm{f}_{\textrm{DFSM}}$.
Once constructed, these models have been used primarily for open-loop optimal control studies.
Methods from surrogate-assisted optimization have been used to refine the DFSM closer to the identified optima.
In this article, we present an approach to constructing a DFSM using principles from systems identification.
We construct this model using the input-output time series data.
% In many cases, the underlying model cannot be directly evaluated to construct the DFSM.
% In this article, we present an approach that uses techniques from systems identification to construct the DFSM.

A key decision that will affect the model quality identified through systems identification techniques is the model type.
A variety of model types have been explored for systems identification approaches, such as time-domain vs.~frequency-domain, continuous-time vs.~discrete-time, linear vs.~nonlinear, etc.
Here, the DFSM is desired to be a continuous-time model.
Once this has been fixed, the next step is to investigate whether a linear or nonlinear system provides more accurate results.
A good starting point to investigate the model type is a LTI system.
It is also straightforward to ensure that the identified linear model is stable.

The system response of wind turbines changes over their operating regions from the cut-in to cut-out flow speeds.
Therefore, a single LTI model would be insufficient to capture the response of wind turbines over their entire operating region, motivating the need for nonlinear models. 
However, ensuring the stability of nonlinear models identified through system identification approaches can be difficult.
The dynamics and response of wind turbines change over their operating region chiefly as a function of the wind speed.
This characteristic of wind turbines has allowed researchers to use linear parameter-varying (LPV) models to construct surrogates~\cite{Sundarrajan2023, Martin2017}.
LPV systems are a type of linear time-varying system where the system matrices are continuous functions of certain parameters.
Typically, the LPV models for wind systems are constructed using a first-order Taylor series approximation of the underlying system model, with the wind speed being the main parameter.
Many articles have explored the construction of LPV models using systems identification principles~\cite{Zhao2012LPV, Casella2008LPV, Toth2010}.
%Few studies investigate the use of systems identification methods to construct LPV models.
The approaches used to develop LPV models in these studies rely on limiting assumptions of the system and cannot be generalized to other systems.
In this article, we present a novel yet simple approach to constructing LPV models of wind turbine systems using time series data.

%----------------
% \clearpage
\xsection{Model Construction}\label{sec:model-construct}

In this section, we outline the different steps involved in constructing the DFSM.
The process starts with defining the key input, controls, and output quantities that we seek to model.
Then, the time series data for these quantities are generated by simulating the system.
Then, using this data, the DFSM is constructed.

\subsection{Modeling Quantities}\label{subsec:model-quantiites}

As shown in Figs.~\ref{fig:workflow} and \ref{fig:dfsm-constrcut-usage}, the three main sets of variables used to construct the low-fidelity models are the primary inputs, controls, and outputs.
The system under consideration is the IEA-15 MW turbine with a semisubmersible platform introduced in Refs.~\cite{Gaertner2020, Allen2020}.
The primary input variables for this system are the wind speed ($w$) and the wave elevation ($\eta$).
As mentioned in Sec.~\ref{subsec:controls}, the generator torque ($\tau_g$) and the blade pitch ($\beta$) are the main control variables for both these turbines.
These two sets of variables are combined and represented as:
\begin{align}
    \label{eq:controls}
    \bm{u} = [w,\tau_g,\beta,\eta]^\mathrm{T}
\end{align}

\noindent where this combined vector contains both kinds of inputs described in Sec.~\ref{subsec:workflow-elements} for simplicity of a single input notation.

Some key outputs that are considered for both these systems are, the generator speed ($\omega_g$), platform surge~($\chi_s$), platform pitch ($\theta_p$), tower base moment in the fore-aft direction ($M_{t,y}$), generator power ($P$), tower-top rotational acceleration in the fore-aft direction ($\ddot{x}_t$), and the tower-top displacement ($\delta_{tt}$).
Of the output quantities listed above, $\omega_g$, $P$, and $\ddot{x}_t$ are feedback variables that must be provided to the controller.
Therefore, to get accurate control responses, the DFSM must capture these variables accurately.
Additionally, quantities such as $M_{t,y}$ and $P$ are used to calculate key performance metrics for these systems, such as the damage equivalent load (DEL) and the annual energy production (AEP), respectively.
\begin{align}
    \label{eq:outputs}
    \bm{y} = [\chi_s,\Theta_p, \delta_{tt},\omega_g, P,M_{t,y},\ddot{x}_t]^\mathrm{T}
\end{align}

\noindent All these quantities are time-varying, but this is not shown explicitly for conciseness.

\subsection{Data Generation and Organization}\label{subsec:data-gen}

Once these quantities have been defined, the FOWT model is simulated using the WEIS toolbox available in Ref.~\cite{WEIS}, using load cases from DLC 1.6.
These load cases are primarily categorized by their average wind speed $(\bar{w})$.
Since the environmental conditions under which a wind turbine must operate are stochastic, evaluating a single set of load cases is often not sufficient to rigorously characterize the performance of the turbine.
Therefore, multiple `seeds' of load cases are generated to simulate the turbine.
Suppose we consider a set of $n_w$ different load cases, each with $n_s$ number of different seeds, the corresponding $\bar{\bm{w}}$ matrix can be represented as follows to obtain a total of $n_s \times n_w$ simulations:
\begin{align}
\label{eq:wind-speed}
\bar{\bm{w}} = \begin{bmatrix}
\bar{w}_{1,1} & \bar{w}_{1,2} & \dots & \bar{w}_{1,n_w}\\ 
\bar{w}_{2,1}  &\bar{w}_{2,2} & \dots & \bar{w}_{2,n_w}\\
\vdots & \vdots & \ddots & \vdots\\
\bar{w}_{n_s,1} &\bar{w}_{n_s,2} & \dots & \bar{w}_{n_s,n_w}
\end{bmatrix}
\end{align}

\noindent The wind turbine is then simulated using OpenFAST for these different load cases to obtain the corresponding combined inputs ($\bm{U}$) database and outputs $(\bm{Y})$ database.
The sizes of these databases are ($n_u \times n_t$) and $(n_y \times n_t)$ respectively, where $n_u$ is the number of combined input/control variables in Eq.~(\ref{eq:controls}), $n_y$ is the number of output variables in Eq.~(\ref{eq:outputs}), and $n_t$ is the number of time points in the simulation.
These quantities are then organized as follows:
% \begin{minipage}{0.44\linewidth}
% \begin{equation}
% \label{eq:U}
% \bm{\mathcal{U}} = \begin{bmatrix}
%     \bm{U}_{1,1}, & \bm{U}_{2,1}, & \dots & \bm{U}_{n_w,1}\\ 
%     \bm{U}_{1,2},  &\bm{U}_{2,2}, & \dots & \bm{U}_{n_w,2}\\
%     \vdots & \vdots & \ddots & \vdots\\
%     \bm{U}_{1,n_s}, &\bm{U}_{2,n_s},0 & \dots & \bm{U}_{n_w,n_s}
% \end{bmatrix}
% \end{equation}
% \end{minipage}
% \hfill
% \begin{minipage}{0.44\linewidth}
% \begin{equation}
% \label{eq:Y}
% \bm{\mathcal{Y}} = \begin{bmatrix}
%     \bm{Y}_{1,1}, & \bm{Y}_{2,1}, & \dots & \bm{Y}_{n_w,1}\\ 
%     \bm{Y}_{1,2},  &\bm{Y}_{2,2}, & \dots & \bm{Y}_{n_w,2}\\
%     \vdots & \vdots & \ddots & \vdots\\
%     \bm{Y}_{1,n_s}, &\bm{Y}_{2,n_s}, & \dots & \bm{Y}_{n_w,n_s}
% \end{bmatrix}
% \end{equation}
% \end{minipage}
%%
%
\begin{gather}
%\label{eq:U-Y}
%\label{eq:U}
\bm{\mathcal{U}} = \begin{bmatrix}
\bm{U}_{1,1} & \bm{U}_{1,2} & \dots & \bm{U}_{1,n_w}\\ 
\bm{U}_{2,1}  &\bm{U}_{2,2} & \dots & \bm{U}_{2,n_w}\\
\vdots & \vdots & \ddots & \vdots\\
\bm{U}_{n_s,1} &\bm{U}_{n_s,2} & \dots & \bm{U}_{n_s,n_w}
\end{bmatrix} \quad
\label{eq:Y}
\bm{\mathcal{Y}} = \begin{bmatrix}
\bm{Y}_{1,1} & \bm{Y}_{1,2} & \dots & \bm{Y}_{1,n_w}\\ 
\bm{Y}_{2,1}  &\bm{Y}_{2,2} & \dots & \bm{Y}_{2,n_w}\\
\vdots & \vdots & \ddots & \vdots\\
\bm{Y}_{n_s,1} &\bm{Y}_{n_s,2} & \dots & \bm{Y}_{n_s,n_w}
\end{bmatrix}
\end{gather}

% \begin{subequations}
% \label{eq:U-Y}
%     \begin{gather}
%         \label{eq:U}
%    \bm{\mathcal{U}} = \begin{bmatrix}
%         \bm{U}_{1,1}, & \bm{U}_{2,1}, & \dots & \bm{U}_{n_w,1}\\ 
%         \bm{U}_{1,2},  &\bm{U}_{2,2}, & \dots & \bm{U}_{n_w,2}\\
%         \vdots & \vdots & \ddots & \vdots\\
%         \bm{U}_{1,n_s}, &\bm{U}_{2,n_s}, & \dots & \bm{U}_{n_w,n_s}
%     \end{bmatrix} \\
%     \label{eq:Y}
%     \bm{\mathcal{Y}} = \begin{bmatrix}
%         \bm{Y}_{1,1}, & \bm{Y}_{2,1}, & \dots & \bm{Y}_{n_w,1}\\ 
%         \bm{Y}_{1,2},  &\bm{Y}_{2,2}, & \dots & \bm{Y}_{n_w,2}\\
%         \vdots & \vdots & \ddots & \vdots\\
%         \bm{Y}_{1,n_s}, &\bm{Y}_{2,n_s}, & \dots & \bm{Y}_{n_w,n_s}
%     \end{bmatrix}
%     \end{gather}
% \end{subequations}

\noindent These matrices correspond to the inputs, controls, and outputs databases shown in Figs.~\ref{fig:workflow}, \ref{fig:surrogate-approaches}, and \ref{fig:dfsm-constrcut-usage}. 

%\vspace{-0.5in}

\subsection{Extracting State Derivative Information}
\label{sec:poly-approx}

In this article, we assume that the selected states are available from the outputs $\bm{y}\in\mathbb{R}^{n_y}$, and the model does not have any other internal states, such that $\bm{\xi} \subset \bm{y}$.
The states considered in this study are $\chi_s,\Theta_p$,$\delta_{tt}$,$\omega_g$, and their first time derivatives.
The corresponding states and state derivatives are:
\begin{align}
    \label{eq:states}
    \bm{\xi} &= [\chi_s,\Theta_p, \delta_{tt},\omega_g,\dot{\chi}_s,\dot{\Theta}_p, \dot{\delta}_{tt},\dot{\omega}_g]^\mathrm{T}\\
    \dot{\bm{\xi}}& =  [\dot{\chi}_s,\dot{\Theta}_p, \dot{\delta}_{tt},\dot{\omega}_g,\ddot{\chi}_s,\ddot{\Theta}_p, \ddot{\delta}_{tt},\ddot{\omega}_g]^\mathrm{T}
\end{align}

When the direct evaluation of $\bm{f}$ is not possible, the state derivative information can be indirectly obtained from the simulated state trajectories.
A continuous polynomial approximation of sampled signals is available in many tools.
% Programming languages like \texttt{MATLAB} and \texttt{Python} have tools to construct a polynomial approximation of given signals.
With at least a $C^1$ approximation, approximate first-order derivatives can be obtained (and higher-order derivatives as well, depending on the method used).
% It is also possible to evaluate the first and second derivatives of the approximations.
Here, a cubic spline polynomial approximation is used to construct continuous $\bm{\xi}(t)$, and then the exact polynomial derivatives are found for  $\dot{\bm{\xi}}(t)$.
Once $\bm{\xi}$ and $\bm{t}$ are available, the functions \texttt{spline}~\cite{spline-mat} and \texttt{fnder}, available in the curve fitting toolbox in \texttt{MATLAB}, and the class \texttt{CubicSpline}~\cite{spline-py} from \texttt{SciPy} can be used to construct the approximation and evaluate the derivatives. 
This approach has been validated using a two-link robotic arm example in Ref.~\cite{Sundarrajan2023}.
The states and state derivatives are then organized as follows:

% \begin{minipage}{0.44\linewidth}
% \begin{equation}
% \label{eq:X}
% \bm{\Xi} = \begin{bmatrix}
% \bm{\xi}_{1,1} & \bm{\xi}_{2,1} & \dots & \bm{\xi}_{n_w,1}\\ 
% \bm{\xi}_{1,2} & \bm{\xi}_{2,2} & \dots & \bm{\xi}_{n_w,2}\\
% \vdots & \vdots & \ddots & \vdots\\
% \bm{\xi}_{1,n_s} & \bm{\xi}_{2,n_s} & \dots & \bm{\xi}_{n_w,n_s}
% \end{bmatrix}
% \end{equation}
% \end{minipage}
% \hfill
% \begin{minipage}{0.44\linewidth}
% \begin{equation}
% \label{eq:X-dot}
% \dot{\bm{\Xi}} = \begin{bmatrix}
% \dot{\bm{\xi}}_{1,1} & \dot{\bm{\xi}}_{2,1} & \dots & \dot{\bm{\xi}}_{n_w,1}\\ 
% \dot{\bm{\xi}}_{1,2} & \dot{\bm{\xi}}_{2,2} & \dots & \dot{\bm{\xi}}_{n_w,2}\\
% \vdots & \vdots & \ddots & \vdots\\
% \dot{\bm{\xi}}_{1,n_s} & \dot{\bm{\xi}}_{2,n_s} & \dots & \dot{\bm{\xi}}_{n_w,n_s}
% \end{bmatrix}
% \end{equation}
% \end{minipage}
\begin{gather}
%\label{eq:X}
\bm{\Xi} = \begin{bmatrix}
\bm{\xi}_{1,1} & \bm{\xi}_{1,2} & \dots & \bm{\xi}_{1,n_w}\\ 
\bm{\xi}_{2,1}  &\bm{\xi}_{2,2} & \dots & \bm{\xi}_{2,n_w}\\
\vdots & \vdots & \ddots & \vdots\\
\bm{\xi}_{n_s,1} &\bm{\xi}_{n_s,2} & \dots & \bm{\xi}_{n_s,n_w}
\end{bmatrix} \quad
\label{eq:X-dot}
\dot{\bm{\Xi}} = \begin{bmatrix}
\dot{\bm{\xi}}_{1,1} & \dot{\bm{\xi}}_{1,2}, & \dots & \dot{\bm{\xi}}_{1,n_w}\\ 
\dot{\bm{\xi}}_{2,1} &\dot{\bm{\xi}}_{2,2} & \dots & \dot{\bm{\xi}}_{1,n_w}\\
\vdots & \vdots & \ddots & \vdots\\
\dot{\bm{\xi}}_{n_s,1} &\dot{\bm{\xi}}_{n_s,2} & \dots & \dot{\bm{\xi}}_{n_s,n_w}
\end{bmatrix}
\end{gather}

% \begin{subequations}
% \begin{gather}
% \label{eq:X}
% \bm{\Xi} = \begin{bmatrix}
% \bm{\xi}_{1,1}, & \bm{\xi}_{2,1}, & \dots & \bm{\xi}_{n_w,1}\\ 
% \bm{\xi}_{1,2},  &\bm{\xi}_{2,2}, & \dots & \bm{\xi}_{n_w,2}\\
% \vdots & \vdots & \ddots & \vdots\\
% \bm{\xi}_{1,n_s}, &\bm{\xi}_{2,n_s}, & \dots & \bm{\xi}_{n_w,n_s}
% \end{bmatrix} \\
% \label{eq:X-dot}
% \dot{\bm{\Xi}} = \begin{bmatrix}
% \dot{\bm{\xi}}_{1,1}, & \dot{\bm{\xi}}_{2,1}, & \dots & \dot{\bm{\xi}}_{n_w,1}\\ 
% \dot{\bm{\xi}}_{1,2}, &\dot{\bm{\xi}}_{2,2}, & \dots & \dot{\bm{\xi}}_{n_w,2}\\
% \vdots & \vdots & \ddots & \vdots\\
% \dot{\bm{\xi}}_{1,n_s}, &\dot{\bm{\xi}}_{2,n_s}, & \dots & \dot{\bm{\xi}}_{n_w,n_s}
% \end{bmatrix}
% \end{gather}
% \end{subequations}

\subsection{Model Construction}\label{sec:DFSM}

The goal of this study is to construct a low-fidelity model that can approximate the key states and outputs of wind turbine systems.
We first assume a nonlinear state-space model with the following model structure:
\begin{subequations}
\label{eq:low-fid}
\begin{align}
\frac{d\bm{\xi}}{dt} = \dot{\bm{\xi}} &= \bm{f}(\bm{\xi},\bm{u}) \approx \bm{f}_{\textrm{LF}}(\bm{\xi},\bm{u})\\
\bm{y} &= \bm{g}(\bm{\xi},\bm{u}) \approx \bm{g}_{\textrm{LF}}(\bm{\xi},\bm{u})
\end{align}
\end{subequations}

We then further assume $\bm{f}_{\textrm{LF}}(\cdot)$ and $\bm{g}_{\textrm{LF}}(\cdot)$ as a LPV state-space models with the following structure:
\begin{subequations}
\label{eq:DFSM-lpv}
\begin{align}
\dot{\bm{\xi}} \approx \bm{f}_{\textrm{LF}}(\bm{\xi},\bm{u}) &= \bm{A}(w)\bm{\xi} + \bm{B}(w)\bm{u}\\
\bm{y} \approx \bm{g}_{\textrm{LF}}(\bm{\xi},\bm{u}) &= \bm{C}(w)\bm{\xi} + \bm{D}(w)\bm{u}
\end{align}
\end{subequations}
%\noindent where $v_{\cdot}$ can be wind speed or current speed depending on the system.

A commonly used approach to construct LPV models is to identify the individual state-space matrices $\bm{\Sigma}_{i} = (\bm{A}_{i},\bm{B}_{i},\bm{C}_{i},\bm{D}_{i})$ corresponding to different values of the system parameter $\bar{w}_i$ where $i\in [1,\dots,n_w]$, and interpolate over them to obtain a continuous representation as shown in Eq.~(\ref{eq:DFSM-lpv}).
We formulate and solve a set of optimization problems to identify the model parameters associated with the system matrices.  
For each value of the parameter $\bar{w}_i$, two different optimization studies are formulated and solved to get the corresponding $\bm{\Sigma}_{i}$, first to get the system matrices $(\bm{A}_{i},\bm{B}_{i})$ corresponding to $\bm{f}_{\textrm{LF}}$, and the second to get the system matrices $(\bm{C}_{i},\bm{D}_{i})$ corresponding to $\bm{g}_{\textrm{LF}}$.

% new paragraph
The primary use case for the model constructed in Eq.~(\ref{eq:DFSM-lpv}) is for time-domain simulations.
For the linear system $\bm{\Sigma}_{i}$ to be stable when simulated, the real parts of the eigenvalues of the state matrix $\bm{A}_{\bar{w}_i}$ must be negative.
The objective for optimization problem solved to obtain the model parameters corresponding to $(\bm{A}_{i},\bm{B}_{i})$ is to minimize the error between the actual state derivatives $(\dot{\bm{\xi}}_{i})$ and the predicted state derivatives $(\hat{\dot{\bm{\xi}}}_{i})$, and has a constraint to ensure that real parts of the eigenvalues are negative.
The optimization problem to identify the model parameters corresponding to $\bm{f}_{\textrm{LF}}$ can be formulated as:
\begin{subequations}
\label{eq:O1}
\begin{align}
% {\bm{\Theta}^{*}_{AB,\bar{w}_i}} = \textrm{arg}\min ||\dot{\bm{\xi}}_{\bar{w}_i} - \hat{\dot{\bm{\xi}}}_{\bar{w}_i}||_2 \\
\text{changing:} \quad & \bm{\theta}_{AB,i} = [\bm{\theta}_{A,i},\bm{\theta}_{B,i}] \\
\text{minimize:} \quad & ||\dot{\bm{\xi}}_{i} - \hat{\dot{\bm{\xi}}}_{i}||_2 \\
\text{subject to:} \quad & \textrm{max}(\textrm{real}(\bm{\lambda}(\bm{\theta}_{A,i})))  < 0 \\
%\text{where:} \quad & \bm{\Theta}_{AB,\bar{w}_i} = [\bm{\Theta}_{A,\bar{w}_i},\bm{\Theta}_{B,\bar{w}_i}]\\
\text{where:} \quad& \bm{A}_{i} = \textrm{mat}(\bm{\theta}_{A,i}) \label{eq:mat1} \\
& \bm{B}_{i} = \textrm{mat}(\bm{\theta}_{B,i}) \label{eq:mat2} \\
& \hat{\dot{\bm{\xi}}}_{i} = \bm{A}_{i}\bm{\xi}_{i} + \bm{B}_{i}\bm{u}_{i}
\end{align}
\end{subequations}

\noindent where $\bm{\lambda}$ are the eigenvalues of $\bm{A}_{i}$, the quantities $(\dot{\bm{\xi}}_{i},\bm{\xi}_{i},\bm{u}_{i})$ are available from the respective columns of the data bases $(\dot{\bm{\Xi}}, \bm{\Xi}, \mathcal{\bm{U}})$ shown in Eqs.~(\ref{eq:Y}) and (\ref{eq:X-dot}) respectively, and `$\textrm{mat}$' in Eqs.~(\ref{eq:mat1}) and (\ref{eq:mat2}) refers to the conversion of a vector into a matrix.
A hybrid optimization strategy is used to solve this problem, where a genetic algorithm (GA) is used to obtain a good estimate for $\bm{\theta}_{AB,i}$. 
A gradient-based solver is used to find the optimal point $\bm{\theta}^{*}_{AB,i}$ starting from the estimate obtained by the GA.

% new paragraph
Similarly, the objective for the optimization problem solved to obtain $(\bm{C}_{i},\bm{D}_{i})$ is to minimize the error between actual outputs $(\bm{y}_{i})$ and the predicted outputs $(\hat{\bm{y}}_{i})$.
The optimization problem to identify the model parameters corresponding to $\bm{g}_{\textrm{LF}}$ can be formulated as:
\begin{subequations}
    \label{eq:O2}
    \begin{align}
    \text{changing:} \quad & \bm{\theta}_{CD,i} = [\bm{\theta}_{C,i},\bm{\theta}_{D,i}] \\
        \text{minimize:} \quad & ||\bm{y}_{i} - \hat{\bm{y}}_{i}||_2^2 \\
        \text{where:}  \quad & \hat{\bm{y}}_{i} = \bm{C}_{i}\bm{\xi}_{i} + \bm{D}_{i}\bm{u}_{i}
    \end{align}
\end{subequations}

\noindent which is a standard linear least-squares estimation problem.

% \noindent The optimization problem in Eq.~(\ref{eq:O2}) can be reformulated as a least-squares estimation problem:
% \begin{subequations}
% \begin{align}
% \label{eq:multi-fidelity-dfsm}
%  \bm{I} &= [\bm{\xi},\bm{u}]^\mathrm{T}\\%\begin{bmatrix}
% % \bm{\xi}\\ \bm{u}
% % \end{bmatrix}\\
% \bm{L} &= [\bm{C},\bm{D}]\\
% \bm{y} \approx \hat{\bm{g}}_{\textrm{LF}}(\bm{\xi},\bm{u}) &=  \bm{L} \bm{I}\\
% \bm{L} &= \bm{I}^\mathrm{T} (\bm{I} \bm{I}^\mathrm{T})^{-1} \bm{y} \label{eq:linsolve}
% \end{align} 
% \end{subequations}

% Tools such as \texttt{linfit} can be used to solve \xdrh{Eq.~(\ref{eq:linsolve})} efficiently.
These optimization problems are formulated and solved to obtain the complete set of state-space matrices $\bm{\Sigma}_i$ for different values of $w_i$. % \in \bm{w}$.

\begin{figure}%[t]
%\centering
%\begin{subfigure}[t]{\columnwidth}
\centering
\noindent\includegraphics[scale=1]{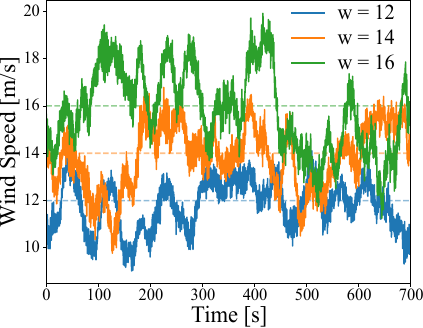}
%\caption{Wind speed trajectory.}
%\label{fig:w_traj} 
%\end{subfigure}%
%---------------------------------------------------
%\begin{subfigure}[t]{\columnwidth}
%\centering
%\noindent\includegraphics[scale=1]{input/plots/w_speed/ws_hist.pdf}
%\caption{Histogram plot of wind speeds.}
%\label{fig:w_hist} 
%\end{subfigure}%
%------------------------------------------------
%-----------------------------------------------------------------
\caption{Wind speed trajectory for three different load cases with $\bar{\bm{w}} = [12,14,16]$ m/s.}
\label{fig:w_speed}
\end{figure}

Some useful properties of FOWT dynamics can be leveraged to efficiently solve the optimization problem formulated to solve Eq.~(\ref{eq:O1}).
In Fig.~\ref{fig:w_speed}, three wind speed trajectories for $\bar{\bm{w}} = [12,14,16]$ [m/s] are plotted.% along with their histograms in Fig.~\ref{fig:w_hist}.
The wind speed trajectory for $\bar{w} = 14$~[m/s] isn't limited to this value, but also contains simulation results where the wind speed is $12$ [m/s] and $16$ [m/s].
The dynamic response of the FOWT depends heavily on the wind speed.
Suppose a state-space model $\bm{\Sigma}$ is constructed to predict $\dot{\bm{\xi}}$ and $\bm{y}$ for a simulation with $\bar{w} = 14$~[m/s].
The model is also trained to be accurate for the parts of the simulation where the wind speed is 12 [m/s] and 16 [m/s].
Therefore, instead of using a GA to identify good starting points for $\bm{\theta}_{AB,12}$ and $\bm{\theta}_{AB,16}$, the optimal point $\bm{\theta}^{*}_{AB,14}$ can be provided as a feasible starting point to the gradient-based solver to obtain $\bm{\theta}^{*}_{AB,12}$ and $\bm{\theta}^{*}_{AB,16}$.
Subsequently, $\bm{\theta}^{*}_{AB,16}$ is used as the starting point for $\bm{\theta}^{*}_{AB,18}$, and $\bm{\theta}^{*}_{AB,12}$ is used as the starting point for $\bm{\theta}^{*}_{AB,10}$, and so on.
An interior-point algorithm is used as the gradient-based solver to leverage this property further.

\xsection{Model Results and Comparisons}\label{sec:validation}

The approach outlined in the previous section is used to construct the DFSM for the FOWT system.
All the OpenFAST simulations in this article are carried out using a high-performance computing cluster with 2$\times$ Intel Xeon Gold 6148 CPU with 40 cores per node and 192 GB of RAM.
All the low-fidelity models used in this article are constructed and simulated using a desktop workstation with an AMD 3970X CPU with 32 cores and 128 GB of RAM, with an NVIDIA GeForce GTX 1650 GPU with 4 GB of VRAM.

\begin{figure}[t]
\centering
\noindent\includegraphics[scale=1]{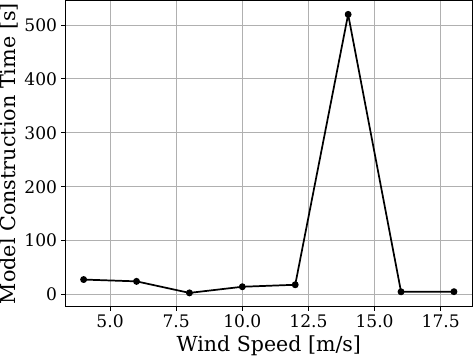}
\caption{Model construction time.}
\label{fig:mc_time} 
\end{figure}%

A total of $n_{s} = 10$ different simulations are obtained for each wind speed.
Of these, $n_{\text{train}} = 5$ seeds are used to train the model, and the rest are used for validation.
The model construction time, along with the set of wind speeds considered for these studies, are shown in Fig.~\ref{fig:mc_time}.
The model construction process starts from $\bar{w} = 14$ [m/s].
A hybrid optimization scheme is used to identify the model parameters for the $\bm{\Sigma}_{i}$ corresponding to this $\bar{w}_i$.
From Fig.~\ref{fig:mc_time}, it can be seen that the time taken to solve Eq.~(\ref{eq:DFSM-lpv}) for this wind speed is 460 seconds.
However, using the approach outlined previously, the solution time for the succeeding wind speeds is significantly reduced, as the parameters identified for $\bar{w} = 14$ [m/s] are used as the starting point for $\bar{w} = 12$ and $\bar{w} = 16$ [m/s].
For example, the construction time for these two succeeding wind speeds is 2.6 seconds.

We would also like to compare the performance of the low-fidelity modeling approach proposed here to low-fidelity models developed using other approaches in the literature, like subspace state-space system identification (n4sid) and models built using LSTM networks.
Using both these alternate approaches results in a model with discrete dynamics.

% \footnote{Pronounced as `enforce it.'} 

\subsection{Subspace State-Space System Identification}
\label{subsec:n4sid}

We utilize the n4sid approach to estimate a discrete-time linear shift-invariant system model using input/output time series data with the following structure:
\begin{subequations}
\label{eq:n4sid}
\begin{align}
\bm{x}_{k+1} &= \bm{A}_{\textrm{d}}\bm{x}_k + \bm{B}_{\textrm{d}}\bm{u}_k \\
\bm{y}_k &= \bm{C}_{\textrm{d}}\bm{x}_k
\end{align}
\end{subequations}

\noindent where the subscript `$\textrm{d}$' denotes a discrete system model.
We use the systems identification toolbox in \texttt{MATLAB} to identify this model.
In addition to the input/output time series data, the model order $n_x$ must be provided by the user.
This selection is identified through a sensitivity study, where we vary $n_x$ for different values and evaluate the response of the identified model.
We test the model response for stability and accuracy.
Through this process, a model with $n_x = 6$ was identified to predict the system response accurately and provide a stable response.
For this system, it is important to use a subspace identification algorithm that can use input/output data obtained using closed-loop simulations~\cite{Jansson2005, VandenHof1998}.
For this reason, we use the `SSARX' algorithm, introduced in Ref.~\cite{Jansson2005} to identify the model in Eq.~(\ref{eq:n4sid}). 

\subsection{Long Short-Term Memory Networks}
\label{subsec:LSTM}

In parallel, we train an LSTM network to predict the outputs for the current time step~($\bm{y}_k$), given the inputs at the current time step~($\bm{u}_k$), and the outputs from the previous time step~$\bm{y}_{k-1}$.
At each timestep, $\bm{u}_k$ is evaluated by using~$\bm{y}_{k-1}$ as the feedback.
The structure of this model is as follows:
\begin{align}
    \label{eq:LSTM}
    \bm{y}_k = \bm{f}_{\textrm{LSTM}}(\bm{y}_{k-1},\bm{u}_k)
\end{align}

The network model contains an LSTM layer with 10 LSTM cells and a fully connected layer.
The number of epochs used to train the model is 10.
We use the implementation in the \texttt{tensorflow}~\cite{tensorflow2015-whitepaper} package to construct and train the model.

\begin{figure}[t]
    \centering
    \includegraphics[width=1\columnwidth]{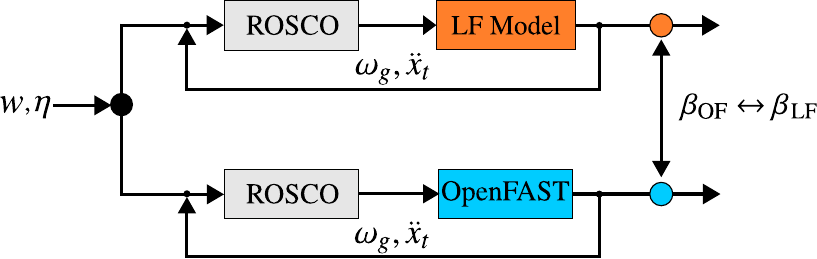}
    \caption{Closed-loop validation test. The low-fidelity models and OpenFAST are used as `plants' in a closed-loop system to provide the main feedback variables of generator speed ($\omega_g$) and tower-top acceleration ($\ddot{x}_t$). For the same wind speed ($w$) and wave elevation ($\eta$) trajectories, the resulting blade pitch ($\beta$) signals are compared.}
    \label{fig:cl-validation}
\end{figure}

\subsection{Closed-Loop Simulation Validation Results}

To validate the models constructed in this section, we carry out the closed-loop validation test as shown in Fig.~\ref{fig:cl-validation}.
The low-fidelity models are used as the `plant' in the closed-loop system and provide the key feedback variables, namely $\bm{\omega}_g$ and $\ddot{\bm{x}}_t$, to the ROSCO controller, which generates the corresponding values for $\tau_g$ and $\beta$.
The resulting control signals obtained using the low-fidelity models are then compared to OpenFAST, which is the high-fidelity model.

% All the OpenFAST simulations in this article are carried out using a high-performance computing cluster with 2$\times$ Intel Xeon Gold 6148 CPU with 40 cores per node and 192 GB of RAM.
% All the low-fidelity models used in this article are constructed and simulated using a desktop workstation with an AMD 3970X CPU with 32 cores and 128 GB of RAM, with an NVIDIA GeForce GTX 1650 GPU with 4 GB of VRAM.

We use the process outlined previously to construct the DFSM.
We use the same dataset to train the n4sid and LSTM models.
Five different seeds are used for training these models.
We train the n4sid and LSTM models on simulations with $\bar{w} = 14$ [m/s].
To obtain accurate predictions with the LSTM model, simulations with $\bar{\bm{w}} = [12,16]$~[m/s] were included as part of the training data.
We utilize the GPU in the desktop workstation for training and evaluating the LSTM model.
The time taken to obtain the training data, in terms of CPU hours, is $5~(n_s) \times 7~(n_w)\times 0.33~\text{hours} = 11.6~\text{hours}$.
Once the data is available, it takes 509 seconds to train the DFSM, 71 seconds to obtain the n4sid model, and 129 seconds to train the LSTM model.
Once trained, we simulate all three models for ten different test cases with $\bar{w} = 14$ [m/s].
For each test case, a 700-second simulation is carried out, but the first 100 seconds of the simulation are discarded to remove the initial transient response.

\begin{figure*}[t]
\centering
\includegraphics[scale=1]{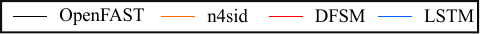}\\
\vspace{2mm}
%-----------------------------------------------------------
\begin{subfigure}[t]{0.33\textwidth}
\centering
\noindent\includegraphics[scale=1]{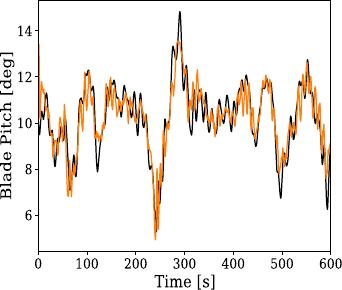}
\caption{n4sid model response.}
\label{fig:BP-n4sid} 
\end{subfigure}%
%---------------------------------------------------
\begin{subfigure}[t]{0.33\textwidth}
\centering
\noindent\includegraphics[scale=1]{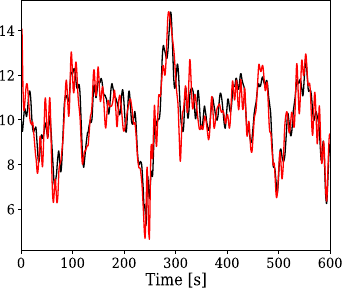}
\caption{DFSM model response.}
\label{fig:BP-DFSM} 
\end{subfigure}%
% %-----------------------------------------------------------
\begin{subfigure}[t]{0.33\textwidth}
\centering
\noindent\includegraphics[scale=1]{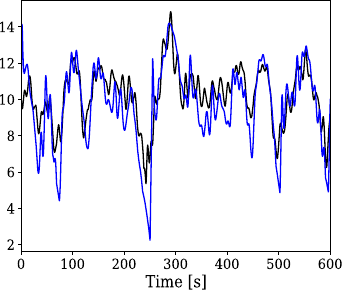}
\caption{LSTM model response.}
\label{fig:BP-LSTM} 
\end{subfigure}%
%-----------------------------------------------------------------
\caption{Comparison of the blade pitch ($\beta$) response obtained using the three different low-fidelity models and OpenFAST as `plants' in a closed-loop simulation.}
\label{fig:val-results}
\end{figure*}

\begin{figure}[t]
\centering
\noindent\includegraphics[scale=0.99]{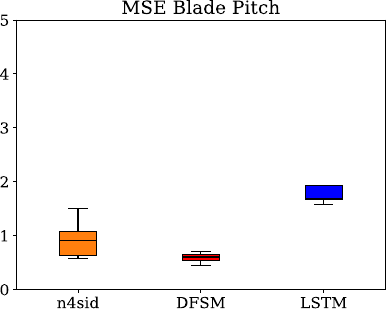}
\caption{MSE vs.~simulation time for all three models.}
\label{fig:BP-MSE} 
\end{figure}

% new paragraph
The blade pitch signal obtained using all three models for one of the test load cases is presented in Fig.~\ref{fig:val-results}.
Additionally, Fig.~\ref{fig:BP-MSE} shows the box plot of the mean square error (MSE) between  OpenFAST and the three different approaches for all ten test load cases.
The average simulation time using the n4sid model is around 2.5 seconds, the average simulation time using the DFSM is 30 seconds, and the average simulation time using the LSTM model is around 70 seconds.
The average simulation time using OpenFAST is nearly 15 minutes.
All three low-fidelity modeling approaches are significantly faster to simulate than OpenFAST.
Although the n4sid model is the least expensive in terms of simulation time.
Conversely, the LSTM model has the highest average simulation time, but the error and the variance in the response are high.
Two of the ten test cases fail for the LSTM model, which does not happen for either the n4sid or DFSM models.
In contrast, the DFSM achieves a good tradeoff between simulation time, model accuracy, and response variance compared to the n4sid and LSTM models.
This can be seen in Fig.~\ref{fig:BP-DFSM}.
Similar tests were carried out for other key output signals such as $\omega_g,M_{t,y},\Theta_p$ as well, and the results follow the trends seen for $\beta$.

% new paragraph
In theory, it should be possible to construct an LPV model using the n4sid approach, similar to the DFSM approach.
However, the choice of $n_x$ that results in an accurate model response differs for different wind speeds.
This makes it hard to obtain an LPV model using the n4sid approaches.
Some additional properties of subspace identification methods make them unsuitable for the task at hand.
The system states ($\bm{x}$) shown in Eq.~\ref{eq:n4sid} and identified by the subspace identification methods are mathematical quantities and do not have any physical meaning.
However, engineers know a priori that certain quantities like $\omega_g$ and $\theta_p$ are system states in physics-based models based on first principles.
However, using the n4sid approach, it is not straightforward to model these quantities as the system states and might be preferred in the model.
%With respect to wind/marine turbines, the state quantities are often known, but it can be difficult to incorporate them as states in the identified model.
% Additionally, n4sid approaches identify discrete-time systems that must then be converted to a continuous-time model\xdrh{, when needed}.
Unlike n4sid models, training a single LSTM network to predict the response for multiple wind speeds is straightforward.
However, using an LSTM network for iterative predictions can be inefficient.
In the study outlined previously, the n4sid model and DFSM are simulated with a timestep of $\Delta_t = 0.01$ [s], while the timestep used for the LSTM model is $\Delta_t = 0.5$ [s].
Even with a larger $\Delta_t$, simulating the system using the LSTM model takes significantly longer.
Additionally, it can be hard to guarantee the stability of the LSTM model.
%---------------
% \clearpage

\xsection{Use Cases}\label{sec:usecases}

In this section, we explore two different use cases for the low-fidelity model, namely closed-loop control simulations and design optimization studies for closed-loop controllers.

\subsection{Closed-Loop Control Simulations}\label{subsec:closed-loop-simulations}

A primary use case for the DFSM constructed in this study is for closed-loop control simulations.
We show how the low-fidelity model can be used to approximate the model response for these systems.
This use case is similar to the validation study discussed in the previous section, but instead of simulating the system for a single wind speed, we simulate the FOWT for multiple wind speeds between its cut-in and cut-out speeds.
In addition to comparing the time series signals obtained using the DFSM and OpenFAST models, we investigate the power spectral density (PSD) plots of some key quantities, the mean value and range of key quantities, and some aggregated quantities like the damage equivalent load (DEL) and the annual energy production (AEP).
A brief description of both of these quantities is given below.

% \begin{figure}[b!]
% \centering
% \noindent\includegraphics[scale=0.99]{input/plots/CL-simulation-MT/w_prob.pdf}
% \caption{Weibull PDF used for the FOWT system.}
% \label{fig:w_prob} 
% \end{figure}%

\subsubsection{Damage equivalent load (DEL):}~%
The DEL metric is often used to approximate the cumulative fatigue damage incurred over a given time span to a single equivalent load that would cause the same damage.
In this article, we are primarily interested in estimating the cumulative fatigue damage incurred at the tower's base. 
The time series signal of tower base moment $M_{t,y}$ is used to calculate the tower base DEL.
The rainflow counting method and the Palmgren-Miner rule are used to estimate the DEL from the $M_{t,y}$ signal.
The reader is referred to the following publications for a detailed explanation of how the DEL is calculated~\cite{Freebury2000, Miao2022}.
The DEL is calculated for the different test cases, and the total DEL is obtained as:
\begin{align}
    \textrm{DEL}_t = \int_w \textrm{DEL}(\bar{w})f_p (\bar{w}) \mathrm{d}w
\end{align}

\noindent where $\textrm{DEL}(\bar{w})$ is DEL for a load case, and $f_p(\bar{w})$ is the probability of $\bar{w}$ obtained from the probability distribution.% shown in Fig.~\ref{fig:w_prob}.

\begin{figure*}[p]
\centering
\includegraphics[scale=1]{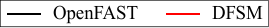}\\
\vspace{1mm}
%-----------------------------------------------------------
\begin{subfigure}[t]{0.33\textwidth}
\centering
\noindent\includegraphics[scale=1]{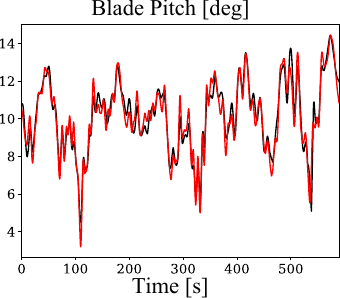}
\caption{Blade pitch ($\beta$).}
\label{fig:BP-comp} 
\end{subfigure}%
%---------------------------------------------------
\begin{subfigure}[t]{0.33\textwidth}
\centering
\noindent\includegraphics[scale=1]{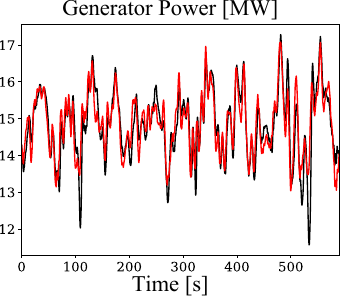}
\caption{Generator power ($P$).}
\label{fig:GS-comp} 
\end{subfigure}%
% %-----------------------------------------------------------
\begin{subfigure}[t]{0.33\textwidth}
\centering
\noindent\includegraphics[scale=1]{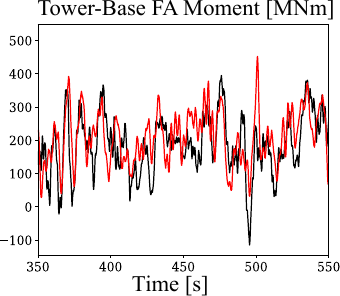}
\caption{Tower-base moment ($M_{t,y}$).}
\label{fig:Myt-comp} 
\end{subfigure}%
%-----------------------------------------------------------------
\caption{Comparison of the time series of key quantities obtained using the DFSM and OpenFAST.}
\label{fig:comp}
\end{figure*}

\begin{figure*}[p]
\centering
\includegraphics[scale=1]{input/plots/CL-simulation-MT/legend-comp.pdf}\\
\vspace{1mm}
%-----------------------------------------------------------
\begin{subfigure}[t]{0.33\textwidth}
\centering
\noindent\includegraphics[scale=1]{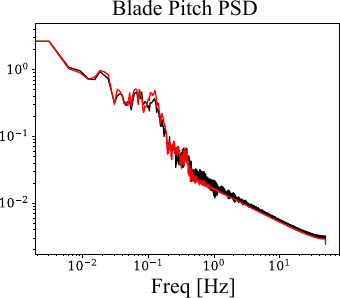}
\caption{Blade pitch ($\beta$).}
\label{fig:BP-PSD-comp} 
\end{subfigure}%
%---------------------------------------------------
\begin{subfigure}[t]{0.33\textwidth}
\centering
\noindent\includegraphics[scale=1]{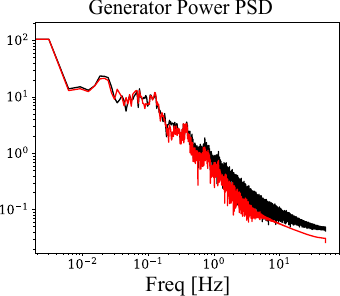}
\caption{Generator power ($P$).}
\label{fig:GS-PSD-comp} 
\end{subfigure}%
% %-----------------------------------------------------------
\begin{subfigure}[t]{0.33\textwidth}
\centering
\noindent\includegraphics[scale=1]{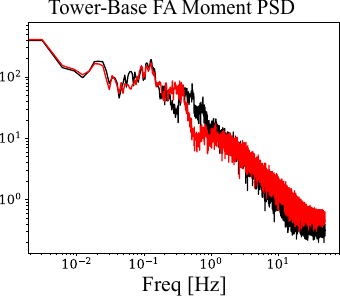}
\caption{Tower-base moment ($M_{t,y}$).}
\label{fig:Myt-PSD-comp} 
\end{subfigure}%
%-----------------------------------------------------------------
\caption{Comparison of the power spectral density (PSD) of the quantities and time series shown in Fig.~\ref{fig:comp}.}
\label{fig:PSD-comp}
\end{figure*}

\begin{figure*}[p]
\centering
\includegraphics[scale=1]{input/plots/CL-simulation-MT/legend-comp.pdf}\\
\vspace{1mm}
%-----------------------------------------------------------
\begin{subfigure}[t]{0.33\textwidth}
\centering
\noindent\includegraphics[scale=1]{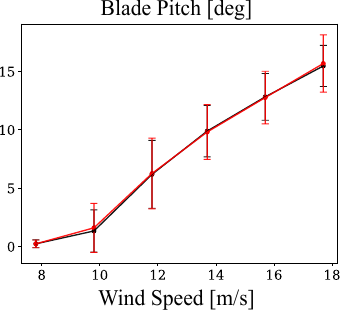}
\caption{Blade pitch $(\beta)$.}
\label{fig:BP-range} 
\end{subfigure}%
%---------------------------------------------------
\begin{subfigure}[t]{0.33\textwidth}
\centering
\noindent\includegraphics[scale=1]{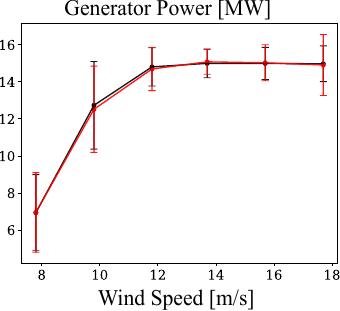}
\caption{Generator power ($P$).}
\label{fig:GP-range} 
\end{subfigure}%
% %-----------------------------------------------------------
\begin{subfigure}[t]{0.33\textwidth}
\centering
\noindent\includegraphics[scale=1]{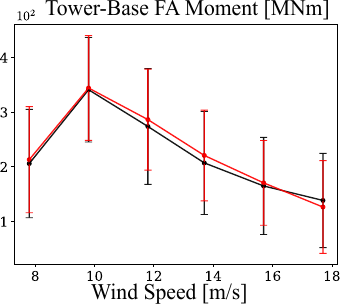}
\caption{Tower-base moment ($M_{t,y}$).}
\label{fig:Myt-range} 
\end{subfigure}%
%-----------------------------------------------------------------
\caption{Comparison of the mean and range over all 60 test load cases simulated using DFSM and OpenFAST for the three quantities.}
\label{fig:range}
\end{figure*}

%\input{input/plots/CL-simulation-MT/CL-DEL-AEP}

% \subsubsection{Annual energy production (AEP):} 
% The AEP is the total energy produced by the turbine in a year of operation.
% It is calculated using the average generator power ($\bar{P}$) for each load case as follows:
% \begin{align}\label{eq:AEP}
%     \textrm{AEP} = 24 \times 365 \times \int_w \bar{P}(\bar{w})f_p (\bar{w}) dw
% \end{align}

We test the system for $n_w = 6$ different load cases, with average wind speed ranging from $\bar{\bm{w}} = [8,10,\cdots,16,18]$ [m/s], with $n_s = 10$ for each $\bar{w}$.
We simulate the system for these load cases using the DFSM and OpenFAST models, and compare the time series signals, PSD plots, and the mean and range of the time series signals from the simulation results.
For the sake of conciseness, we plot the time series and PSD results of a single load case in the rated region.
For this load case, we focus on three main quantities, namely $\beta$, $P$, and $M_{t,y}$.
The time series comparison between the DFSM and OpenFAST responses for these quantities for a single load case is shown in Fig.~\ref{fig:comp}.
The time taken to simulate the load case using the DFSM is around 25 seconds, whereas it takes nearly 20 minutes to simulate the OpenFAST model.
As can be seen from Figs.~\ref{fig:BP-comp} and \ref{fig:GS-comp}, the DFSM can predict the response of $\beta$ and $P$ with a high degree of accuracy.
Although not shown in the article, the efficacy of the DFSM in predicting $\omega_g$ and $\ddot{x}_t$ can be inferred from  Fig.~\ref{fig:BP-comp}.
As mentioned in Sec.~\ref{subsec:controls}, these two quantities are the main feedback variables to calculate $\beta$.
However, as can be seen from Fig.~\ref{fig:Myt-comp}, the DFSM cannot predict the highly nonlinear signal $M_{t,y}$ with the same degree of accuracy.
Although the DFSM can capture the mean trends for this signal, the linear approximation is not sufficient to capture the entirety of the signal accurately.

% new paragraph
In addition to the time series signals, in Fig.~\ref{fig:PSD-comp}, we show the PSD plots for the signals shown in Fig.~\ref{fig:comp}.
The key frequencies for these systems are highlighted in the PSD plot.
The 3P frequency, or the thrice-per-rotation frequency, is the frequency at which the rotors cross the tower.
The 3P frequency is critical for these three quantities, and a peak occurs around this frequency in the corresponding PSD plots of these signals. 
Additionally, the wave profile acting on the platform affects the $M_{t,y}$ signal as well.
From Fig.~\ref{fig:Myt-PSD-comp}, it can be seen that the DFSM response can capture this as well, even though there is a large discrepancy between the time series signal predicted by the DFSM and OpenFAST.
Although there is a slight difference between the power density values in this range, the peaks occur around the same key frequencies.
Since the goal is to use the DFSM rapidly in early-stage design studies, some accuracy loss might be expected as long as the DFSM can identify the right trade-offs.

These results show the performance of the DFSM for a single load case.
Figure~\ref{fig:range} shows the mean values and range for the time series signals of the three quantities over all $10~(n_s) \times 6~(n_w) = 60$ test load cases considered for this study.
Similar to the results shown in Fig.~\ref{fig:comp}, the DFSM accurately predicts the mean and range for quantities like $\beta$ and $P$.
The DFSM can predict the mean for the $M_{t,y}$ signal consistently, but it underpredicts the range for this signal.
This affects the quantile metrics, like the DEL calculated using the response of DFSM.

\subsection{Controller Optimization Study}\label{subsec:control-opt}

\begin{figure*}[t]
\centering
\begin{subfigure}[t]{0.5\textwidth}
\centering
\noindent\includegraphics[scale=1]{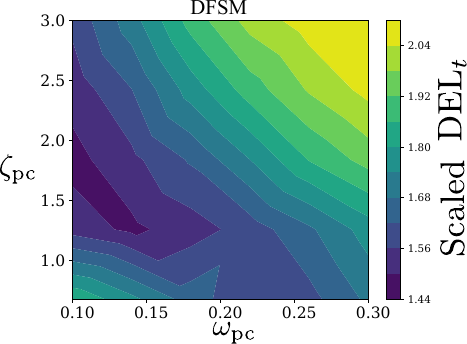}
\caption{\DEL{} calculated using DFSM simulations.}
\label{fig:DEL-DFSM} 
\end{subfigure}%
%---------------------------------------------------
\begin{subfigure}[t]{0.5\textwidth}
\centering
\noindent\includegraphics[scale=1]{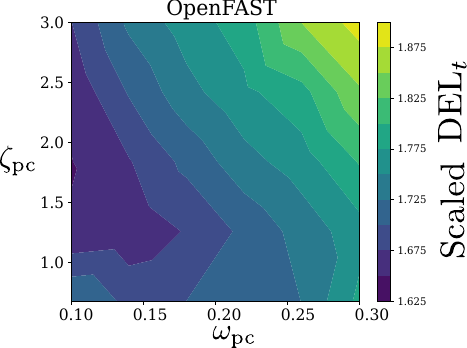}
\caption{\DEL{} calculated using OpenFAST simulations.}
\label{fig:DEL-OF} 
\end{subfigure}%
%------------------------------------------------
%-----------------------------------------------------------------
\caption{Comparison of the \DEL{} vs.~$\bm{x}_c$ calculated using DFSM and OpenFAST simulations.}
\label{fig:DOE-DEL}
\end{figure*}

%\input{input/plots/DOE/DEL_scaled}

% new paragraph
The second use case for the DFSM modeling approach presented in this article is to explore the use of these low-fidelity models in design optimization studies.
The blade pitch controller is crucial in minimizing the fatigue load and maximizing the power capture.
For this reason, many controller optimization studies for wind and marine turbines focus on the blade pitch controller~\cite{Sundarrajan2024, Zalkind2022}.
The design variables considered here are the natural frequency and damping ratio of the blade pitch controller:
\begin{align}
    \bm{x}_c = [\omega_{\textrm{PC}},\zeta_{\textrm{PC}}]^\mathrm{T}
\end{align}

\noindent where the subscript `$\textrm{PC}$' stands for pitch controller.
We simulate the system using the DFSM and OpenFAST for five key wind speeds with $\bar{\bm{w}} = [10,12,14,16,18]$.
We use $n_s = 5$ for each wind speed.
These wind speeds are chosen because they are in the rated region, where the blade pitch controller is active and highly weighted in the Weibull PDF we use for this system.
%, as can be seen in Fig.~\ref{fig:w_prob}.
Because of this, DEL values calculated for these load cases are weighted higher than others.

% new paragraph
The validation studies carried out for the DFSM previously assume that the $\bm{x}_c$ does not change.
But to be able to use the DFSM for controller optimization studies, it is necessary to test the DFSM for different values of $\bm{x}_c$.
We conduct a design of experiments (DOE) study, using a full-factorial sampling scheme to generate $n = 36$ different samples between its lower and upper bounds.
We then simulate the system using the DFSM and OpenFAST models at these sample points for the generated load cases, and plot the resulting design space for the \DEL{} in Fig.~\ref{fig:DOE-DEL}, with Fig.~\ref{fig:DEL-DFSM} showing the design space obtained with the DFSM, and Fig.~\ref{fig:DEL-OF} showing the design space obtained using OpenFAST.

It can be clearly seen from Fig.~\ref{fig:DOE-DEL} that even though the DFSM underpredicts the \DEL{} value for this system, the shape and the trends in the design space are similar to what is obtained using OpenFAST.
The average simulation time for a single load case using the DFSM is around 25 seconds, whereas it takes nearly 20 minutes to simulate the same load case using OpenFAST.
Therefore, the total computational cost in terms of CPU hours required to simulate all 30 different load cases at all 25 sampling points is approximately $36~(n) \times 6~(n_s) \times 5~(n_w)\times15/60~(\text{hours}) = 270$ (hours) using OpenFAST but only around $36~(n) \times 6~(n_s) \times 5~(n_w)\times35/3600~(\text{hours}) = 10.5$ (hours) using the DFSM.
This study shows that the DFSM approach can be effective in predicting the system response for different $\bm{x}_c$ values, even though it was \textit{not} trained on these.

% new paragraph
With the DFSM validated, we formulate and solve the following blade pitch controller optimization study to minimize \DEL{}.
We extend the controller design variables to include variables associated with the floating-feedback controller as well
The complete design optimization problem can be formulated as:
\begin{subequations}
\label{eq:multifid-formulation}
\begin{align}
    \text{changing:}\quad & \bm{x}_c = [\omega_{\textrm{pc}},\zeta_{\textrm{pc}},k_{\beta,\textrm{float}},\omega_{\textrm{ptfm}}]^T\\
    \text{minimize:}\quad & o = \textrm{DEL}_t\\
    \text{subject to:}\quad & \omega_g \leq 1.25\times\omega_{g,\textrm{rated}} \label{eq:constraint}\\
    &\bm{x}_{c,\min} \leq \bm{x}_c \leq \bm{x}_{c,\max}\\
    \text{where:}\quad &\bm{x}_{c,\min} = [0.1,0.6,-40,0.0]^T\\
    &\bm{x}_{c,\max} = [0.3,3.0,0.0,0.4]^T
\end{align}
\end{subequations}
\noindent Where, the constraint in Eq.~(\ref{eq:constraint}) is called the rotor-overspeed constraint to ensure the rotor speed and by extension, the generator speed does not exceed $25$\% of its rated value, with $\omega_{g,\textrm{rated}} = 7.5$ [rpm], for this turbine.

To account for the difference in predictions between the DFSM and OpenFAST, we utilize a multi-fidelity trust-region approach presented in Ref.~\cite{Jasa2022}, which is based on the methods presented in Refs.~\cite{Alexandrov2001, March2012}.
At the core of this approach is an additive corrective function ($e(\cdot)$) to correct the low-fidelity function value:
\begin{align}
    \label{eq:corrective}
    f_{\textrm{high}}(\bm{x}_{c}) \approx f_{\textrm{surrogate}}(\bm{x}_c) =f_{\textrm{low}}(\bm{x}_{c}) + e(\bm{x}_{c})
\end{align}
Once constructed, the surrogate model is used to solve the optimization problem in Prob.~(\ref{eq:multifid-formulation}) iteratively using a trust-region approach.
In this multi-fidelity approach, the DFSM is the low-fidelity model, and OpenFAST is the high-fidelity model. 
Although not shown explicitly, all the elements of Prob.~(\ref{eq:multifid-formulation}) are scaled appropriately.
Readers are referred to Ref.~\cite{Jasa2022} for more details on this approach and the software implementation.
The optimization problem presented in Prob.~(\ref{eq:multifid-formulation}) is solved using the DFSM, OpenFAST, and the multi-fidelity approach.

\begin{table*}[t]
\normalsize
\renewcommand{\arraystretch}{1}
\begin{center}
\caption{Multi-fidelity controller optimization results for Prob.~\ref{eq:multifid-formulation}.}
\label{tab:optparameters-MF}
\begin{tabular}{cccccc}
  \hline\hline
 {Fidelity} & {LF calls} & {HF calls} & {$\bm{x}_c$} & {$\textrm{DEL}_{t,\textrm{HF}}~\text{at}~\bm{x}_c$} & {$\omega_g/\omega_{g,rated}~\text{at}~\bm{x}_c$} \\ 
  \hline
Low-fidelity & 24 & -- & $[0.14,1.65,-07.69,0.36]^T$ & $1.62 \times 10^5$ & 1.23\\ 
Multi-fidelity& 468 & 11 & $[0.15,2.12,-15.37,0.38]^T$ & $1.56 \times 10^5$ & 1.24  \\ 
High-fidelity& -- & 38 & $[0.10,2.92,-14.42,0.36]^T$ & $1.55 \times 10^5$ & 1.23\\
\hline \hline
\end{tabular}
\end{center}
\end{table*}

The results obtained using the three models are presented in Table~\ref{tab:optparameters-MF}.
All three optimization studies were started from the same point for $\bm{x}_{c,\textrm{init}} = [0.18,2.2,-4.9,0.2]^T$.
The \DEL{} value at this point for the high-fidelity function is $1.79 \times 10^5$ [kNm].
It is clear from Table~\ref{tab:optparameters-MF}, that both the DFSM and multi-fidelity approach converge to a point that has a lower value of \DEL{}.
Furthermore, using the multi-fidelity approach, it is possible to identify a point that is closer in terms of objective function value to the OpenFAST response.
The difference between the identified optima for the multi-fidelity and high-fidelity models can be attributed to the trust region-based approach and the optimizer used to solve the trust-region problem.
However, the percentage error between the identified optima is just $0.64$ \%.

%\clearpage
\subsection{Multi-Objective Controller Optimization From a Control Co-Design Perspective}
\label{subsec:MOMF}

In this section, we formulate and solve a multi-objective controller optimization problem. 
For this case study, we use a slightly modified version of the IEA-15 MW turbine with a semisubmersible platform. 
This model has a different tower and platform design compared to the systems discussed previously; consequently, the \DEL{} values reported for this system differ slightly from those presented earlier. 
We consider load cases corresponding to five different mean wind speeds, $\bm{w}_{\textrm{avg}} = [10, 12, 14, 16, 18]$, using six seeds for each wind speed. 
The weighted-objective method is employed to solve this multi-objective problem.
Maximizing performance metrics such as power production often conflicts with the goal of minimizing fatigue damage. 
This trade-off is explored in detail in Ref.~\cite{Sundarrajan2025c} between the power quality, measured as the standard deviation of the generator speed~($\omega_{g,\textrm{std}}$), while \DEL{} is used to measure the fatigue damage. 
Ideally, when designing the turbine, the goal is to minimize both these metrics. 
However, these objectives are conflicting. 
In addition to the controller, the design of the platform also has an effect on these quantities.
It is therefore important to study this trade-off from a control co-design perspective as well.
In this study, we focus on the column spacing, denoted as $\bm{x}_{\xrm{p}}$, as the primary plant design variable. 
In addition to $\bm{x}_{\xrm{p}}$, we vary the controller parameters $\bm{x}_c$ associated with the blade-pitch controller introduced previously.

We adopt a nested control co-design (CCD) formulation to carry out this study.  
In the outer loop, the column spacing is varied, while in the inner loop, we solve the multi-objective controller optimization problem for the corresponding platform configuration. 
Since the outer loop considers only a single design variable, we perform a sensitivity study by varying the column spacing between its upper and lower bounds. 
Specifically, the column spacing is evaluated at five discrete values:
\begin{align}
    \bm{x}_{\textrm{p}} = [35,42,51,57,65]
\end{align}
\noindent In the inner loop, we utilize a weighted sum approach to solve the multi-objective optimization problem.

We formulate the following multi-objective optimization problem:
\begin{subequations}
\label{eq:multiobj-formulation1}
\begin{align}
\text{changing:}\quad & \bm{x} = [\bm{x}_{\textrm{p}},\bm{x}_c]^T \\
        \text{minimize:}\quad & \bm{o} = [\textrm{DEL}_t,\omega_{g,\textrm{std}}]\\
        \text{subject to:} \quad & \bm{x}_{c,\min} \leq \bm{x}_c \leq \bm{x}_{c,\max} \\
    &\bm{x}_{c,\min} \leq \bm{x}_c \leq \bm{x}_{c,\max}\\
    \text{where:}\quad &\bm{x}_{c,\min} = [0.1,0.6,-40,0.0]^T\\
    &\bm{x}_{c,\max} = [0.3,3.0,0.0,0.4]^T
\end{align}
\end{subequations}

\subsubsection{Updating the DFSM:}~%

Changing the column spacing affects the dynamic response of the turbine, particularly quantities that capture floating motion and loads, such as $\Theta_p$, $\chi$, and $M_{t,y}$. 
%For the same load case, systems with different values of $x_{\textrm{p}}$ exhibit distinct responses for these quantities. 
Therefore, as the column spacing varies, the DFSM must be updated accordingly. 
The approach used previously to construct the DFSM---running multiple 600-second load cases for each wind speed---could be applied here as well. 
However, performing multiple OpenFAST simulations for each iteration of the outer loop can be computationally expensive. 
To mitigate this, we modify the DFSM construction process for the nested CCD approach.

Previously, model identification began with a hybrid-optimization approach to identify an LTI model for a specific wind speed, which then served as the starting point for successive wind speeds. 
We adopt the same principle when updating the DFSM for a new value of $x_{\textrm{p}}$. 
We start with the DFSM identified for the nominal column spacing $x_{\textrm{p}} = 51$~[m], whose construction began at $w = 14$~[m/s]. 
For a platform with $x_{\textrm{p}} = 42$~[m], we also start at $w = 14$~[m/s], but instead of the hybrid-optimization approach, we use the LTI model identified for $x_{\textrm{p}} = 51$~[m] as the initial guess for a gradient-based optimizer. 
Additionally, we reduce the simulation length from 600 seconds to 300 seconds during model identification. 
Once the LTI model is identified for $w = 14$~m/s at $x_{\textrm{p}} = 42$~[m], it serves as the starting point for $w = 12$~[m/s], and for a turbine with for $x_{\textrm{p}} = 35$~[m], and so on. 
%Similarly, the LTI model for $x_{\textrm{p}} = 42$~[m] is used as the starting point when constructing the DFSM for $x_{\textrm{p}} = 35$~[m]. 
This strategy significantly reduces the computational cost of constructing the DFSM for different values of $\bm{x}_{\xrm{cs}}$.
Validation results for the updated models can be found in Ref.~\cite{Sundarrajan2025c}.

\subsubsection{Results:}
\begin{figure*}[t]
    \centering
    \includegraphics[scale = 1.1]{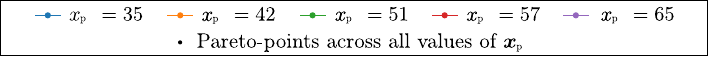}\\
    \vspace{2mm}
    \begin{subfigure}{0.5\textwidth}
    \includegraphics[scale = 1]{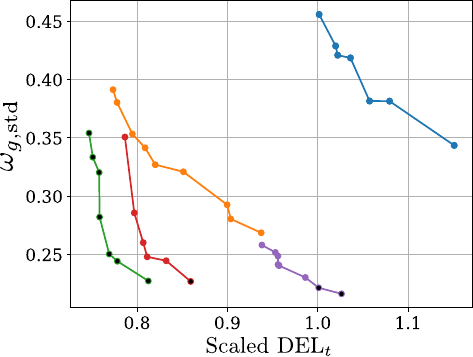}
    \caption{Individual Pareto fronts for all the different values of $\bm{x}_{\textrm{p}}$.}
    \label{fig:pareto-ccd}
    \end{subfigure}%
    \begin{subfigure}{0.5\textwidth}
    \includegraphics[scale = 1]{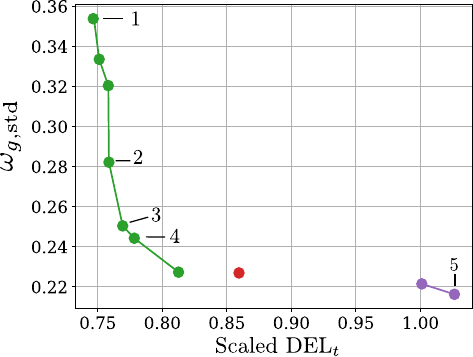}
    \caption{Pareto front obtained using the CCD study across all $\bm{x}_{\textrm{p}}$.}
    \label{fig:pareto-focus}
    \end{subfigure}
    \caption{Pareto front between \DEL{} vs. $\omega_{g,\textrm{std}}$ obtained for different values of column spacing for the IEA-15 MW turbine with a semisubmersible platform.}
\end{figure*}

~%
With the DFSM constructed and validated for different values of $\bm{x}_{\xrm{cs}}$, we carry out the multi-objective controller optimization study using the multi-fidelity approach outlined previously, and report the results for the high-fidelity model. 
The optimization problem is solved for platforms with $\bm{x}_{\textrm{p}} = [42,51,57,65]$~[m], twice for each, starting from different initial points. 
For $x_{\textrm{p}} = 35$~[m], this is not performed, as preliminary studies indicated that both objectives are significantly higher compared to the other platforms. 
The resulting Pareto fronts for these multi-objective optimization problems are shown in Fig.~\ref{fig:pareto-ccd}. 
Across all Pareto fronts, it is evident that \DEL{} and $\omega_{g\text{std}}$ remain conflicting objectives, although the level of conflict varies between different platforms.

% \begin{figure*}
% \centering
% \includegraphics[scale = 1.3]{5/pareto/legend-MO-CCD.pdf}\\
% \vspace{2mm}
% \begin{subfigure}{0.5\textwidth}
%     \centering
%     \includegraphics[scale = 1]{5/pareto/pareto-CCD-4var.pdf}
%     \caption{Pareto-fronts for all values of $\bm{x}_{\textrm{cs}}$.}
%     \label{fig:pareto-ccd-all}
% \end{subfigure}%
% \begin{subfigure}{0.5\textwidth}
%     \centering
%     \includegraphics[scale = 1]{5/pareto/pareto-focus.pdf}
%     \caption{Pareto-front for the CCD study.}
%     \label{fig:pareto-ccd-focus}
% \end{subfigure}
% \caption{Pareto-front between \DEL{} vs. $\omega_{g,\textrm{std}}$ for different values of $\bm{x}_{\textrm{cs}}$ .}
% \label{fig:pareto-ccd}
% \end{figure*}

\begin{figure*}
\centering
\vspace{2mm}
\begin{subfigure}{0.5\textwidth}
    \centering
    \includegraphics[scale = 1]{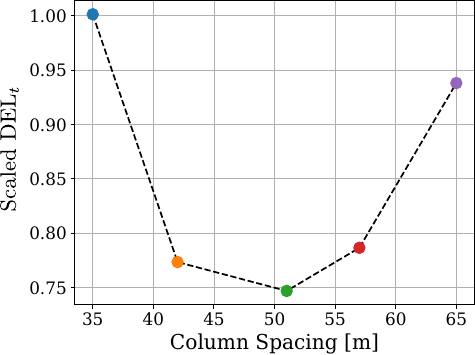}
    \caption{Optimal \DEL{} vs. column spacing.}
    \label{fig:DEL-CS}
\end{subfigure}%
\begin{subfigure}{0.5\textwidth}
    \centering
    \includegraphics[scale = 1]{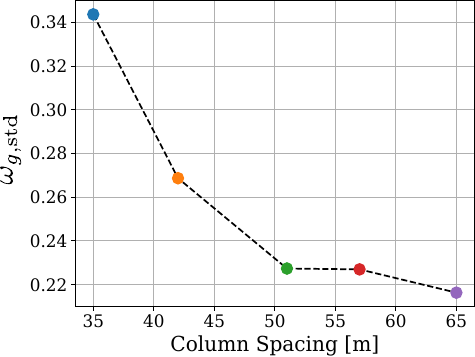}
    \caption{Optimal $\omega_{g,\textrm{std}}$ vs. column spacing.}
    \label{fig:gsstd-CS}
\end{subfigure}
\caption{Comparison of the optimal \DEL{} and $\omega_{g,\textrm{std}}$ vs.~column spacing obtained as part of the multi-objective optimization results shown in Fig.~\ref{fig:pareto-ccd}.}
\label{fig:OBJvsCS}
\end{figure*}
\begin{figure*}[t]
\centering
\includegraphics[scale =1]{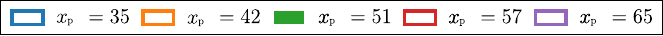}\\
\vspace{2mm}
\begin{subfigure}{0.5\textwidth}
    \centering
    \includegraphics[scale =1]{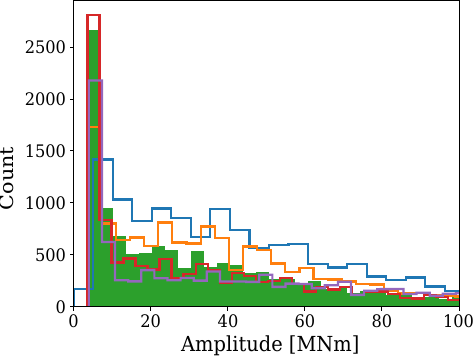}
    \caption{Amplitude and count in the range of 0--100 [MNm].}
    \label{fig:DEL-all}
\end{subfigure}%
%\hspace{0.025\textwidth}%
\begin{subfigure}{0.5\textwidth}
    \centering
    \includegraphics[scale =1]{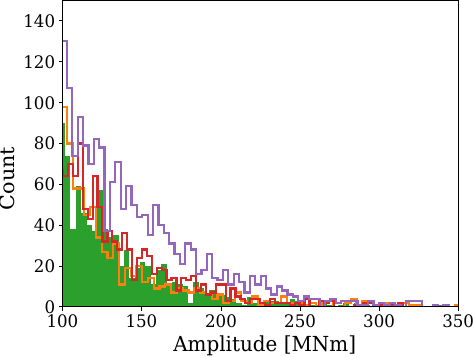}
    \caption{Amplitude and count in the range of 100--350 [MNm].}
    \label{fig:DEL-high-mag}
\end{subfigure}
\caption{The amplitude and count for the different load cycles from all thirty simulations used to calculate the \DEL{} for the different values of $\bm{x}_{\textrm{p}}$.}
\label{fig:DEL-hist}
\end{figure*}

The behavior of the optimal values of \DEL{} and $\omega_{g\text{std}}$ differs as $x_{\textrm{p}}$ is varied. 
\DEL{} exhibits a nonlinear trend with respect to column spacing, whereas $\omega_{g\text{std}}$ monotonically decreases as the column spacing increases. 
The platform with $x_{\textrm{p}} = 35$~[m] has the highest optimal $\omega_{g\text{std}}$, and the platform with $x_{\textrm{p}} = 65$~[m] has the lowest, as shown in Figs.~\ref{fig:DEL-CS} and \ref{fig:gsstd-CS}, respectively.

The trends in Fig.~\ref{fig:DEL-CS} can be explained physically. 
For $x_{\textrm{p}} = 35$~[m], the platform is smaller and more susceptible to large motions, which dominate \DEL{}. 
For $x_{\textrm{p}} = 65$~[m], the platform is heavier and stiffer, resulting in smaller motion amplitudes. 
However, reduced compliance means more aerodynamic thrust is transmitted directly into the tower, increasing base loads despite smaller platform motions.
These trends can be further understood by examining the \DEL{} calculation and corresponding histogram plots. 
Figure~\ref{fig:DEL-hist} shows histograms of load cycle amplitudes from 30 simulations for all values of $\bm{x}_{\xrm{cs}}$, using the controller parameters that yield the lowest \DEL{}. 
Figure~\ref{fig:DEL-all} presents amplitudes below 100~[MNm], while Fig.~\ref{fig:DEL-high-mag} shows amplitudes above 100~[MNm] for $\bm{x}_{\xrm{cs}} = [42,51,57,65]$~[m], with the histogram for $x_{\textrm{p}} = 51$ highlighted.

From Fig.~\ref{fig:DEL-all}, it is evident that the platform with $x_{\textrm{p}} = 35$~[m] has more high-amplitude load cycles, whereas the platform with $x_{\textrm{p}} = 65$~[m] has a higher count of cycles with mean amplitudes between 100--250~[MNm]. 
Consequently, these systems experience higher cumulative damage and \DEL{}. 
Platforms with $x_{\textrm{p}} = [51,57]$~[m] have more load cycles with lower amplitudes, while $x_{\textrm{p}} = 42$~[m] has a higher count of cycles in the 0–100~[MNm] range. 
The key factor influencing \DEL{} is the number of cycles in the 100–250~[MNm] range. 
The platform with $x_{\textrm{p}} = 51$~[m] satisfies both criteria---high count of low-amplitude cycles and low count of high-amplitude cycles---resulting in the lowest \DEL{}.

The combined Pareto front, containing the non-dominated points across all values of $x_{\textrm{p}}$, is shown in Fig.~\ref{fig:pareto-focus}. 
All points from the individual Pareto fronts are part of this combined front. 
This further validates selecting $x_{\textrm{p}} = 51$~[m] as the optimal platform design, as it contributes the highest number of points to the combined Pareto front. 
For this platform, the optimal controller parameters must be sub-selected. 
Consider points labeled 1--5 in Fig.~\ref{fig:pareto-focus}: point 1 has the lowest \DEL{} and highest $\omega_{g,\text{std}}$, while point 5 has the lowest $\omega_{g,\text{std}}$ and highest \DEL{}, illustrating the trade-off between these two objectives.

\begin{table}[t]
\normalsize
\renewcommand{\arraystretch}{1}
\begin{center}
% \caption{\DEL{} and $\omega_{g,\textrm{std}}$ values for the five points labeled in Fig.~\ref{fig:pareto-focus}, where $\Delta \mathrm{DEL}_t = 100\frac{\mathrm{DEL}t-\mathrm{DEL}{t,\min}}{\mathrm{DEL}{t,\min}}$, and $\Delta \omega_{g,\textrm{std}} = $100 $\times \frac{(\omega_{g,\textrm{std}}- \omega_{g,\textrm{std},\min})}{\omega_{g,\textrm{std},\min}}$.}
\caption{\DEL{} and $\omega_{g,\textrm{std}}$ values for the five points labeled in Fig.~\ref{fig:pareto-focus}.}
\label{tab:ccd-focus}
\begin{tabular}{ccccc}
  \hline\hline\\[-13pt]
 {Point} & {\DEL{}} & {$\omega_{g,\textrm{std}}$} & $\Delta \mathrm{DEL}_t$ [\%] & $\Delta \omega_{g,\textrm{std}}$ [\%]  \\ 
  \hline
1 & 0.7469 & 0.3541 & 0.00 & 63.77 \\ 
2& 0.7585 & 0.2821 & 1.56  & 30.46\\ 
3& 0.7692 & 0.2504 & 1.99  & 15.78\\ 
4& 0.7782 & 0.2442 & 4.19 & 12.93\\
5 & 1.0263 & 0.2162& 37.40& 0.00\\
\hline \hline
\end{tabular}
{\footnotesize $\Delta \mathrm{DEL}_t = 100\frac{\mathrm{DEL}t-\mathrm{DEL}{t,\min}}{\mathrm{DEL}{t,\min}}$ \hspace{0.25in} $\Delta \omega_{g,\textrm{std}} = $100 $\times \frac{(\omega_{g,\textrm{std}}- \omega_{g,\textrm{std},\min})}{\omega_{g,\textrm{std},\min}}$}
\end{center}
\end{table}

Comparing point 2 to point 1, we observe a $1.56\%$ increase in \DEL{} but a $30.46\%$ increase in $\omega_{g,\text{std}}$ relative to point 5.  
For point 3, the \DEL{} increases by $1.99\%$ and $\omega_{g,\text{std}}$ increases by $15.78\%$.  
Similarly, comparing point 4 to points 1 and 5, there is a $4.19\%$ increase in \DEL{} and a $12.93\%$ increase in $\omega_{g,\text{std}}$.  
Considering all these points, point 3 achieves a favorable trade-off between the two objectives.  

The corresponding optimal design values are:
\begin{align}
    x_{\textrm{p,opt}} &= 51~\text{[m]} \\
    \bm{x}_{c,\textrm{opt}} &= [0.1931,2.885,-14.464,0.2185]^T
\end{align}

%This study highlights the inherent trade-offs between \DEL{} and $\omega_{g,\text{std}}$ and demonstrates the value of the control co-design (CCD) approach in systematically exploring these trade-offs.

%\clearpage
%---------------
%\clearpage
\xsection{Conclusion}\label{sec:conclusion}

In this article, we investigate an approach to building continuous-time low-fidelity models of complex dynamic systems such as floating wind turbines --- termed a derivative function surrogate model (DFSM).
We proposed an approach to extract the state derivative information from system simulations by constructing polynomial approximations of the states and evaluating the derivatives of these approximations.
With the extracted state derivative information, we presented a novel way to construct a linear-parameter varying state-space model that can predict key system states and outputs, given the controls.
We explored how the DFSM approach compares to different data-driven low-fidelity modeling approaches, like system identification and deep learning, presented in the literature.
The results from these studies show that the DFSM approach is more consistent when it comes to predicting the dynamics of the wind turbine system considered in this study.

The constructed DFSM addresses two main drawbacks when using high-fidelity models, namely for estimating system performance and for use in different types of design optimization studies.
Three different studies are carried out to test the effectiveness of the DFSM in addressing these drawbacks.
The first study is a closed-loop control study, where the DFSM is used as the plant in a closed-loop system. 
The system is simulated for different test cases, and the controls and outputs predicted by the DFSM are compared to the high-fidelity model response.
The results show that the DFSM developed using the approach presented can accurately predict the system dynamics at a fraction of the computational cost and can be used to generate key time series outputs of the controls and outputs.
The computational time to carry out one high-fidelity simulation is nearly 15 minutes, whereas the DFSM can be simulated in 35 seconds.
These time series can be post-processed to obtain key performance metrics for the system.

In the second study, we carried out an optimization study for key controller design variables.
This study started with mapping out the associated design space for key controller variables against a performance metric like the tower-base damage equivalent load (\DEL{}) using both the low-fidelity DFSM model and the high-fidelity OpenFAST model.
%When using the high-fidelity model, high performance computing resources are need to run this study.
%The controller variables are varied over the lower and upper bounds, and the low and high-fidelity models are evaluated at these points to get the key performance metric like the tower-base damage equivalent load (\DEL{}).
%A contour plot is then generated, showing how the \DEL{} changes with the design variables for the low and high fidelity models.
The results show that the DFSM underpredicts the metric, but it can accurately capture the shape and key trends in this design space.
The DFSM offers nearly a 48 times speedup compared to the high-fidelity models for this study, enabling users to carry out this study on a personal laptop computer.
Based on the insight from this study, we opted to use a trust-region-based multi-fidelity optimization framework to effectively use the low- and high-fidelity models to identify the optimal controller variables that minimize the \DEL{}.
The study shows that the approach and the low-fidelity model can be used to identify a point that is close to the high-fidelity minima with fewer high-fidelity function calls.

Finally, in the third study, we tested how amenable the DFSM is to a simple multi-objective control co-design, with the goal of identifying a design that balances conflicting objectives of \DEL{} and power quality, measured as the standard deviation of the generator speed.
The plant variable for this study is simply the column spacing of the floating platform.
This study was designed to showcase the effectiveness of the DFSM even when it needs to be updated for different $\bm{x}_c$.
We present a way to quickly update the DFSM for different plant designs.
We use a nested approach to solve the problem, in which the outer loop updates the plant design serially, and the inner loop carries out the multi-objective optimization study.
Using this approach, we show that the DFSM can be used to identify a point that satisfies the trade-off between these two points.

In addition to the data-driven modeling approaches discussed in this study, other novel approaches must also be tested for their efficacy in modeling the system response.
Models such as deep state space models (DSSM)~\cite{Gedon2021, Pillonetto2025}, an approach combining techniques from systems identification and deep learning, are examples of these approaches that can be used to approximate the system for the use cases discussed in this article.
Additionally, networks such as continuous-time echo state networks (CTESN) or neural ordinary differential equations~(nODE), also offer an approach to model continuous-time systems~\cite{anantharaman2020, Roberts2022, Chen2018}.
Improvements could also be made to the DFSM approach presented in this study.
For example, the DFSM approach presented in this article could also be augmented with other nonlinear data-driven modeling approaches to predict quantities like the tower-base moment accurately.
The efficacy of the DFSM approach in predicting a wider set of states and outputs must also be investigated further.

Finally, future research could focus on investigating approaches to extend the methodology presented here to create low-fidelity models of other systems as well.
Utilizing data from real systems to construct low-fidelity models could also be explored.

%Additionally, the possibility of constructing the DFSM using shorter simulations must also be investigated.

%Although it is possible to extend the DFSM for control co-design studies using an approach as shown in Ref.~\cite{Sundarrajan2023}, it currently has several challenges to overcome.
%The approach presented in Ref.~\cite{Sundarrajan2023} involves sampling the plant design space, constructing a low-fidelity model at each sample point, and interpolating over them to obtain a continuous model.
%This approach can be inefficient for the reasons outlined in Sec.~\ref{subsec:surrogates}.
%Instead, researchers should consider the DFSM approach presented in this study as part of a nested control co-design strategy.
%--------------
%\clearpage
\begin{acknowledgment}
The information, data, or work presented herein was funded by the Advanced Research Projects Agency-Energy (ARPA-E), U.S. Department of Energy, under Award Number DE-AC36-08GO28308.
The authors would like to thank Dan Zalkind, Thanh Toan Tran, and Hannah Ross of NLR, and Saeed Azad of Ohio Northern University for their feedback.
\end{acknowledgment}

%--------------------------
% \clearpage
\renewcommand{\refname}{REFERENCES}
\bibliographystyle{config/asmems4}
\begin{mySmall}
%\nocite{*} % remove later, displays all references
\bibliography{References}
\end{mySmall}

%--------------------------
% \clearpage
% \onecolumn
% \xneed{Will be removed in the final version.}
% % This will be removed in the final version.
% \tableofcontents
% \listoffigures
% \listoftables
% -----------

%---------------
%\clearpage
%\input{input/working.tex}

\end{document}